\documentclass[12pt]{article}
\usepackage{epsfig}

\newcommand{\bea}{\begin{eqnarray}}
\newcommand{\eea}{\end{eqnarray}}
\newcommand{\be}{\begin{equation}}
\newcommand{\ee}{\end{equation}}
\newcommand{\MSbar}{\overline{\rm MS}}
\newcommand{\as}{\alpha_s}
\newcommand{\asMZ}{\alpha_s(M^2_Z)}
\newcommand{\ar}{a_s}

\title{QCD coupling constant at NNLO from DIS data
\author{B.G.~Shaikhatdenov$^a$, A.V.~Kotikov$^a$, V.G.~Krivokhizhin$^a$, and G.~Parente$^b$ \\
$^a$ Joint Institute for Nuclear Research, Russia \\
$^b$ Universidade de Santiago de Compostela, Spain} }

\begin{document}
\maketitle
\abstract{ Deep inelastic scattering data on $F_2$ structure function from the various fixed-target experiments were analyzed in the non-singlet approximation with a next-to-next-to-leading-order accuracy.
The study of high statistics deep inelastic scattering data provided by BCDMS, SLAC, NMC and BFP collaborations was carried out separately for the first one and the rest, followed by a combined analysis done as well. For the coupling constant the following value $\as(M_Z^2) = 0.1167 \pm 0.0021 \mbox{\scriptsize{(total exp.error)}}+\biggl\{
\begin{array}{l} +0.0056 \\ -0.0036 \end{array} ~\mbox{\scriptsize{(theor)}}$ was found, which in this approximation turns out to be slightly less than that obtained at the next-to-leading-order, as was generally anticipated. Ditto the theoretical uncertainties reduced with respect to those obtained in the case of the next-to-leading-order analysis thus confirming earlier observations.} \\

$PACS:~~12.38~Aw,\,Bx,\,Qk$\\


\section{ Introduction }

It goes without saying how it is crucial to know as accurate as possible the parton
distribution functions (PDFs) and the value of the strong coupling constant in order
to be able to make (relatively) solid predictions for various processes studied in a number of experiments. Within this realm, the deep inelastic scattering (DIS) of leptons off hadrons serves to be a cornerstone process to study PDFs which are universal and feed them further to other processes.

Nowadays the accuracy of data for DIS structure functions (SFs) makes it possible to study $Q^2$-dependence of logarithmic QCD-inspired corrections and those of power-like (non-perturbative) nature in a separate way (see for instance~\cite{Beneke} and references therein) which is important for the analysis to be performed according to a well defined scheme.

Until recently a commonly adopted benchmark tool for the analysis happened to be there
at the next-to-leading-order (NLO) level. However there have already appeared papers in which QCD analysis of DIS SFs has been carried out up to the next-to-next-to-leading order (NNLO) (see e.g.~\cite{PKK}-\cite{NNLOBlumlein}  and references therein).

The present paper closely follows the one devoted to the similar study performed at NLO level~\cite{KK2001} with the major difference in that here we deal with the nonsinglet case only because of relative complicacy of the task considered. The singlet part of the analysis (combined with the nonsinglet one) will be accomplished in the near future.
We analyze DIS SF $F_2(x,Q^2)$ with SLAC, NMC, BCDMS and BFP experimental
data involved~\cite{SLAC1}--\cite{BFP} at NNLO of massless perturbative QCD.
This has become possible thanks to the results on both the $\alpha_s^3(Q^2)$ corrections to the splitting functions (the anomalous dimensions of Wilson operators)~\cite{MVV2004} and the corresponding expressions of the complete three-loop coefficient functions for the structure functions $F_2$ and $F_L$~\cite{MVV2005}.

As in our previous paper the function $F_2(x,Q^2)$ is represented as a sum of the leading twist $F_2^{pQCD}(x,Q^2)$ and the twist four terms~\footnote{This form was used in \cite{KK2001,KK2009}, too: Eq. (3.32) in~\cite{KK2009} should be replaced by~(\ref{1.1}).}:
\be
F_2(x,Q^2)=F_2^{pQCD}(x,Q^2)\left(1+\frac{\tilde h_4(x)}{Q^2}\right)\,.
\label{1.1}
\ee
While analysing experimental data various corrections must be taken into account. Here the nuclear effects, target mass corrections, heavy quark threshold corrections and higher twist terms are considered. For details we refer to~\cite{KK2001,KK2009}.

As is known there are at least two ways to perform QCD analysis over DIS data: the first one (see e.g.~\cite{ViMi,fits}) deals
with Dokshitzer-Gribov-Lipatov-Altarelli-Parisi (DGLAP) integro-differential equations~\cite{DGLAP} and let the data be examined directly, whereas the second one involves the SF moments and permits performing an analysis in analytic form as opposed to the former option.
In this work we take on the way in-between these two latter, i.e. analysis is carried out over the moments of SF $F_2^{k}(x,Q^2)$
defined as follows~\footnote{Hereinafter, $k=pQCD,twist$ denotes the twist two approximation with and without target-mass corrections (see, for example, \cite{KK2001}).}
\be
M_n^{pQCD/twist2/\ldots}(Q^2)=\int_0^1 x^{n-2}\,F_2^{pQCD/twist2/\ldots}(x,Q^2)\,dx
\label{1.a}
\ee
and then reconstruct SF for each $Q^2$ by using Jacobi polynomial expansion method \cite{Barker}-\cite{Kri1} (for further details see~\cite{KK2001,KK2009} and section~\ref{sec3}).

%========================================================================================
\section{ A brief theoretical input }

Here we briefly touch on certain aspects of the theoretical part of our analysis.
For a bit detailed account see~\cite{KK2001}.
The twist-two DIS SF can be represented as a sum of two terms:
$
F_2^{twist2}(x,Q^2)= F_2^{NS}(x,Q^2) + F_2^{S}(x,Q^2)\,,
$
the nonsinglet (NS) and singlet (S) parts. At this point let's introduce PDFs, the gluon distribution function $f_G(x,Q^2)$
and the singlet and nonsinglet quark distribution functions
${\bf f}_S(x,Q^2)$ and ${\bf f}_{NS}(x,Q^2)$~\footnote{Unlike the standard case, here PDFs are multiplied by $x$.}:
\bea
{\bf f}_{S}(x,Q^2) &\equiv& \sum_{q}^{f} {\bf f}_{q}(x,Q^2)= V(x,Q^2) + S(x,Q^2)\,, \nonumber \\ \nonumber
{\bf f}_{NS}(x,Q^2) &=& {\bf u}_v(x,Q^2) - {\bf d}_v(x,Q^2)\,,
\eea
where $f$ is the number of quark flavors (${\bf u}$p, ${\bf d}$own,
${\bf s}$trange,$\ldots$), $V(x,Q^2)={\bf u}_v(x,Q^2)+{\bf d}_v(x,Q^2)$ is the distribution of valence quarks and $S(x,Q^2)$ is
a sum of sea parton distributions set equal to each other.

There is a direct relation between SF moments~(\ref{1.a}) and those of PDFs
$$
{\bf f}_{NS}(n,Q^2) ~=~\int_0^1 dx x^{n-2} {\bf f}_{NS}(x,Q^2). %\label{3.aa}
$$
For example, in the nonsinglet case it looks~\cite{Buras}:
\be
M_n^{NS}(Q^2) = R_{NS}(f)\times C_{NS}^{twist2}(n,\ar(Q^2))\times {\bf f}_{NS}(n,Q^2)\,,
\label{3.a}
\ee
with
\be
\ar(Q^2)=\frac{\alpha_s(Q^2)}{4\pi} \label{as}
\ee
and $C_{NS}^{twist2}(n,\ar(Q^2))$ are the Wilson coefficient functions.
The constant $R_{NS}(f)$ depends on the weak and electromagnetic charges and is fixed to be one sixth for $f=4$~\cite{Buras}.

%%%%%%%%%%%%%%%%%%%%%%%
\subsection{Strong coupling constant}

The strong coupling constant is determined from the corresponding solution of the renormalization group equation to an accuracy of ${\cal O}(10^{-5})$ (which is enough for our purposes, also we checked that for higher precision the results get no much better). At NLO level the latter is given by
\bea \label{1.coA}
\frac{1}{\ar^{NLO}(Q^2)} - \frac{1}{\ar^{NLO}(M_Z^2)} +
b_1 \ln{\left[\frac{\ar^{NLO}(Q^2)}{\ar^{NLO}(M_Z^2)}
\frac{(1 + b_1\ar^{NLO}(M_Z^2))}
{(1 + b_1\ar^{NLO}(Q^2))}\right]}
= \beta_0 \ln{\left(\frac{Q^2}{M_Z^2}\right)}\,.
\eea

At NNLO level the strong coupling constant is derived from the following equation:
\bea \label{1.co}
\frac{1}{\ar(Q^2)} - \frac{1}{\ar(M_Z^2)} &+&
b_1 \ln{\left[\frac{\ar(Q^2)}{\ar(M_Z^2)}
\sqrt{\frac{1 + b_1\ar(M_Z^2) + b_2\ar^2(M_Z^2)}
{1 + b_1\ar(Q^2) + b_2\ar^2(Q^2)}}\right]} \\ \nonumber
&+& \left(b_2-\frac{b_1^2}{2}\right)\times
\Bigl(I(Q^2)- I((M_Z^2)\Bigr) = \beta_0 \ln{\left(\frac{Q^2}{M_Z^2}\right)}\,.
\eea
The expression for $I$ looks:
$$
I(Q^2)=\cases{
\displaystyle{\frac{2}{\sqrt{\Delta}}} \arctan{\displaystyle{\frac{b_1+2b_2\ar(Q^2)}{\sqrt{\Delta}}}} &for $f=3,4,5; \Delta>0$,\cr
\displaystyle{\frac{1}{\sqrt{-\Delta}}}\ln{\left[
\frac{b_1+2b_2\ar(Q^2)-\sqrt{-\Delta}}{b_1+2b_2\ar(Q^2)+\sqrt{-\Delta}}
\right]}&for $f=6;\quad\Delta<0$, \cr
}
$$
where $\Delta=4b_2 - b_1^2$ and $b_i=\frac{\beta_i}{\beta_0}$ are read off from the QCD $\beta$-function:
$$
\beta(\ar) ~=~ -\beta_0 \ar^2 - \beta_1 \ar^3 - \beta_2 \ar^4 +\ldots
$$

The equations (\ref{1.coA}) and (\ref{1.co}) allow us to eliminate QCD
parameter $\Lambda_{{\rm QCD}}$ from the analysis. However, sometimes it is appropriate to consider it. %we consider it in our discussions so as to be able to compare with the results of older fits.
The coupling constant $\ar(Q^2)$ is expressed through $\Lambda_{QCD}$
(in $\MSbar$ scheme, where $\Lambda_{{\rm QCD}}=\Lambda_{\MSbar}$) as follows:

at NLO level
\bea
\frac{1}{\ar^{NLO}(Q^2)} +
%\frac{\beta_1}{\beta_0}
b_1 \ln{\left[\frac{\beta_0 \ar^{NLO}(Q^2)}
{1+ b_1 \ar^{NLO}(Q^2)}\right]} =
\beta_0
\ln{\left(\frac{Q^2}{\Lambda^2_{\MSbar,NLO}}\right)}
\label{2.coA},
\eea

and at NNLO level
\bea \label{2.co}
\frac{1}{\ar(Q^2)} &+&
b_1 \ln{\left[\frac{\beta_0\ar(Q^2)}
{\sqrt{1 + b_1\ar(Q^2) + b_2\ar^2(Q^2)}}\right]}  \nonumber \\
&+& \left(b_2-\frac{b_1^2}{2}\right)\cdot(I(Q^2) - I(0)) =
\beta_0 \ln{\left(\frac{Q^2}{\Lambda^2_{\MSbar}}\right)}\,.
\eea

A relation between the constant at the normalization point $\ar(M_Z^2)$ and QCD parameter
$\Lambda_{{\rm QCD}}$ can be obtained from Eqs.~(\ref{2.coA}) and (\ref{2.co}) by substituting $Q^2$ for $M_Z^2$.

Note that sometimes (see, for example,~\cite{PKK}) the equations
\bea
\frac{1}{\ar^{NLO}(Q^2)} + b_1
\ln{\left(\beta_0 \ar^{NLO}(Q^2) \right)} =
\beta_0
\ln{\left(\frac{Q^2}{\Lambda^2_{\MSbar,NLO}}\right)}
\label{3.coA},
\eea
and
\bea
\frac{1}{\ar(Q^2)} + b_1
\ln{\left(\beta_0 \ar(Q^2) \right)} + (b_2-b_1^2)\ar(Q^2)  =
\beta_0
\ln{\left(\frac{Q^2}{\Lambda^2_{\MSbar}}\right)}
\label{3.co},
\eea
are used in the analyses with NLO and NNLO approximations, respectively.
These can be deduced from the basic equation
\bea
\ln{\left(\frac{Q^2}{\Lambda^2_{\MSbar}}\right)}~=~
\int^{\ar(Q^2)}
\frac{db}{\beta(b)},
\label{4.co}
\eea
by expanding an inverse QCD $\beta$-function in RHS of Eq.~(\ref{4.co})
(that is $1/\beta(\ar)$) in powers of $\ar$ up to $O(\ar)$ and $O(\ar^2)$, respectively.
The difference between Eqs.~(\ref{3.coA}),~(\ref{3.co})
and~(\ref{2.coA}),~(\ref{2.co}) can reach ${\cal O}(10^{-3})$ at $Q^2\sim 1$~GeV$^2$ energies.
To avoid uncertainties caused by this approach we use in the analyses a numerical solution (with an accuracy of $10^{-5}$) of Eq.~(\ref{1.co}) instead. Let's mention in this regard that the approximations given in Eqs.~(\ref{3.coA}),~(\ref{3.co}) and~(\ref{2.coA}),~(\ref{2.co}), based on the expansion of inverse powers of $\ln{\left(Q^2/\Lambda^2_{\MSbar}\right)}$ are very popular on the market. They have the following forms:
\bea
\ar^{NLO}(Q^2) = \frac{1}{\beta_0 L_{NLO}} -
\frac{b_1\ln L_{NLO}}{(\beta_0 L_{NLO})^2}
+ {\cal O}((\beta_0 L_{NLO})^{-3})\,,
\label{3.coB}
\eea
and
\bea
\ar(Q^2) = \frac{1}{\beta_0 L} - \frac{b_1\ln L}{(\beta_0 L)^2}
+ \frac{1}{(\beta_0 L)^3} \, \Bigl[b_1^2(\ln^2 L-\ln L-1)+b_2\Bigr]
+ {\cal O}((\beta_0 L)^{-4})\,,
\label{3.coC}
\eea
where $L_{NLO}=\ln(Q^2/\Lambda^2_{NLO})$ and $L=\ln(Q^2/\Lambda^2)$ in the NLO and NNLO approximations, in order.
%An accuracy of these expansions for the evolution of $\ar$
%from ${\cal O}$ (GeV$^2$) to $M_Z^2$ may be as large as $0.001$~\cite{Al2001},
%which is comparable with the experimental uncertainties of the $\as(M_Z^2)$
%value extracted from the data (see our analyses in Sections 4 and 5).
\vspace{0.5cm}
\begin{figure}[!htb] %\label{fig-1}
\unitlength=1mm
\vskip -1.5cm
\begin{picture}(0,100)
\put(0,-5){%
   \psfig{file=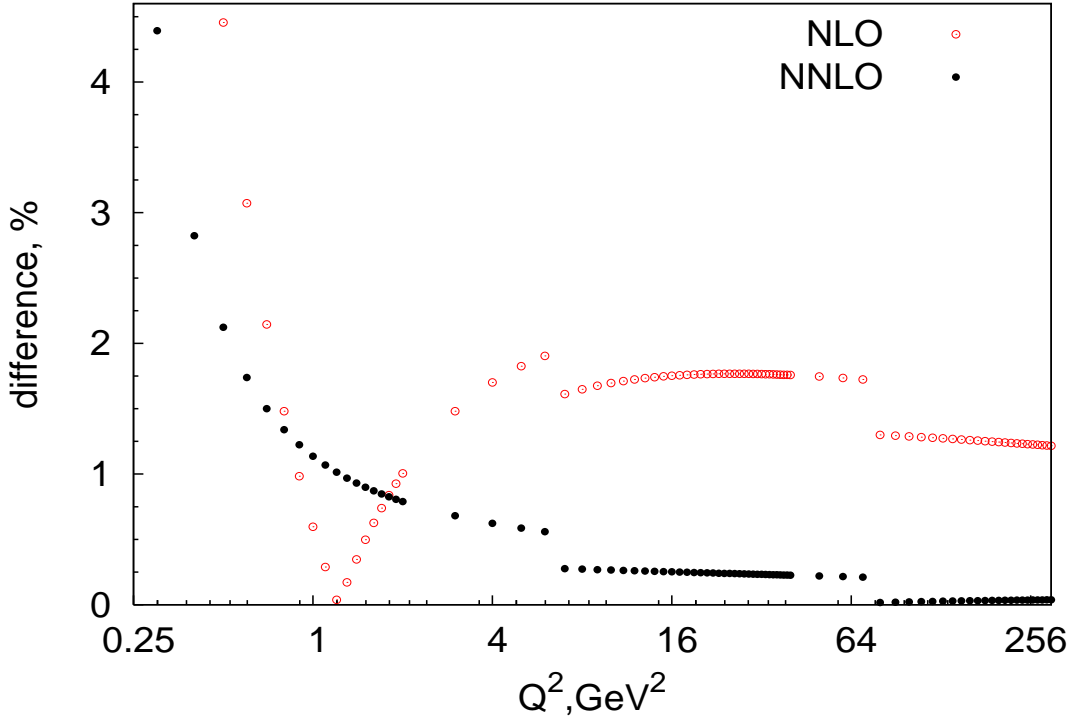,width=150mm,height=100mm}
}
\end{picture}
\vskip 0.2cm
\caption{\it{
Difference between a numeric solution to Eqs.~(\ref{2.coA}) and~(\ref{2.co})
(as precise as ${\cal O}(10^{-5})$) and approximate representations given in~(\ref{3.coB}) and~(\ref{3.coC}) for the strong coupling constant at NLO and NNLO, respectively}.}
\end{figure}
\vspace{0.5cm}

Note that the difference at NNLO level, shown in Fig.~1, between an approximate expression
for $\alpha_s$ sometimes used in the literature (see, e.g.,~\cite{Iran}) and
a solution to Eq.~(\ref{2.co}), hence Eq.~(\ref{1.co}), becomes less than that obtained in NLO~\cite{Melnit} and still of order ${\cal O}(10^{-3})$ (observed
in~\cite{Al2001,Blumlein96}) in the range $Q^2$ spanning in this analysis
(see also discussion in~\cite{KK2001}), which is comparable with the experimental uncertainties of $\as(M_Z^2)$ value obtained from the data (see our analyses in Secs.~4 and~5). Although a utilization of the exact transcendental equation in the NNLO case for deriving the coupling constant appears to be not much of a preference over the approximate expression for the latter within an almost entire scan range in $Q^2$ (as opposed to the NLO case), we still prefer to carry out the analysis with the former option for it is still an exact (to the order considered) equation to be numerically solved to the accuracy desired.

Also note that the results shown in Fig.~1 were obtained with
$\Lambda(f=4) =$ 200 MeV, which is less than those obtained below. Therefore, the actual difference between the exact formul\ae\, given in Eqs.~(\ref{1.coA}),~(\ref{1.co}) and, respectively, their approximations quoted in Eqs.~(\ref{3.coB}),~(\ref{3.coC}) is even a bit more pronounced.

%showshere
A starting point of the evolution is taken at relatively large values $Q_0^2$. There is a number of reasons behind that choice, e.g., fewer heavy quark thresholds have to be crossed to reach a normalization point, a perturbative approach must be applicable at the value of $Q^2_0$. Besides, impact of higher order corrections derived from PDF normalization conditions is the more negligible the higher normalization point is.

%%%%%%%%%%%%%%%%%%%%%%%
\subsection{$Q^2$-dependence of SF moments}

The coefficient functions $C_{NS}^{twist2}(n,\ar(Q^2))$ is further expressed through
the functions $B^j_{NS}(n)$ which are known exactly~\cite{MVV2005,Buras}~\footnote{For the odd $n$ values, the $F_2$ coefficients $B^j_{NS}(n)$ and $Z^j_{NS}(n)$ can be obtained using the analytic continuation~\cite{Kri1,KaKo,KoVe}.}
\be
C_{NS}^{twist2}(n, \ar(Q^2)) = 1 % - \delta^G_i
+ \ar(Q^2) B^{NLO}_{NS}(n)
+ \ar^2(Q^2) B^{NNLO}_{NS}(n) + {\cal O}(\ar^3(Q^2))\,.
\label{1.cf}
\ee

The $Q^2$-evolution of the PDF moments can be calculated within the framework
of perturbative QCD (see e.g.~\cite{Buras,Ynd}):
\be
\frac{{\bf f}_{NS}(n,Q^2)}{{\bf f}_{NS}(n,Q_0^2)}=\left[\frac{\ar(Q^2)}
{\ar(Q^2_0)}\right]^{\frac{\gamma_{NS}^{(0)}(n)}{2\beta_0}}
\times H^{NS}(n,Q^2, Q_0^2)\,. % \nonumber \\
\label{3}
\ee
The function $H^{NS}(n, Q^2, Q^2_0)$ %and $H_j^{\pm}(n, Q^2, Q^2_0)$
up to NNLO may be represented as %~\footnote{There is an option to proceed with calculating by using the exact exponential expressions for both $H^{NS}(n,Q^2,Q^2_0)$ and coefficient functions. We've checked that the difference in results originating from utilizing this alternative is found to be ${\cal O}(10^{-5})$ and can safely be neglected to the accuracy we work with, thus justifying the actual use of the expansion forms, given in the text, in the evaluations throughout.}
%\begin{eqnarray}\label{hnnsA}
%H^{NS}(n, Q^2, Q^2_0) &=& 1 + \left[ \ar(Q^2) - \ar(Q^2_0) \right] Z^{NLO}_{NS}(n) \nonumber \\
%&-& \left[\ar(Q^2)-\ar(Q^2_0)\right]\ar(Q^2_0) [Z^{NLO}_{NS}(n)]^2 \\ \nonumber
%&+& \left[\ar^2(Q^2)-\ar^2(Q^2_0)\right] Z^{NNLO}_{NS}(n)
%+ {\cal O}\left(\ar^3(Q^2)\right)\,,
%\end{eqnarray}
%or
\begin{eqnarray}\label{hnns}
H^{NS}(n, Q^2, Q^2_0) &=& \frac{h^{NS}(n, Q^2)}{h^{NS}(n, Q^2_0)}, \nonumber \\
h^{NS}(n, Q^2)  &=& 1 + \ar(Q^2) Z^{NLO}_{NS}(n) + \ar^2(Q^2) Z^{NNLO}_{NS}(n)
+ {\cal O}\left(\ar^3(Q^2)\right)\,,
\end{eqnarray}
where~\cite{KKPS1}
\begin{eqnarray}
Z^{NLO}_{NS}(n) &=& \frac{1}{2\beta_0} \biggl[ \gamma_{NS}^{(1)}(n) -
\gamma_{NS}^{(0)}(n)\, b_1\biggr]\,, \nonumber \\
Z^{NNLO}_{NS}(n)&=& \frac{1}{4\beta_0}\left[
%2\beta_0 Z^2_{NS}(n) +
\gamma^{(2)}_{NS}(n)
-\gamma^{(1)}_{NS}(n)b_1 + \gamma^{(0)}_{NS}(n)(b^2_1-b_2) \right]
+  \frac{1}{2} Z^2_{NS}(n)
\,.
\label{3.21}
\end{eqnarray}
Here $\gamma_{NS}^{(k)}(n)$ are the factors before $\ar$ in the expansion
with respect to the latter of the anomalous dimensions $\gamma_{NS}(n,\ar)$
(taken in the exact form from~\cite{MVV2004}).
%In the analysis the form given in Eq.~(\ref{hnns}) rather than the one presented in Eq.~(\ref{hnnsA}) is used. The latter is utilized while comparing to the results of our earlier work, in which it was actually used.

%%%%%%%%%%%%%%%%%%%%%%%
\subsection{Factorization $\mu_F$ and renormalization $\mu_R$ scales}

Also, we are set to consider the dependence of results on the factorization $\mu_F$ and
renormalization $\mu_R$ scales, caused by (see, e.g.,~\cite{ViMi, scheme, Blumlein96})
the truncation of a perturbative series while doing the calculus. A modification is
achieved by replacing $\ar$ (defined in Eq.~(\ref{as})) in Eqs.~(\ref{3.a},\ref{3}) with the expressions in which the scales were accounted in the following way: $\mu^2_F = k_F Q^2,\quad\mu^2_R= k_R \mu^2_F = k_R k_F Q^2$.

Then, Eq.~(\ref{3.a}) takes the form:
\be
M_n^{NS}(Q^2) = R_{NS}(f) \times \hat C_{NS}^{twist2}(n, \ar(k_F Q^2))
\times {\bf f}_{NS}(n,k_F Q^2), \nonumber
\ee
and Eq.~(\ref{3}) gets replaced by
\be
\frac{{\bf f}_{NS}(n, k_F Q^2)}{{\bf f}_{NS}(n,k_F Q_0^2)}
=\bigg[\frac{\ar(k_F k_R Q^2)}
{\ar(k_F k_R Q^2_0)}\bigg]^{\gamma_{NS}^{(0)}(n)/2\beta_0}
\times \hat H^{NS}(n,k_F k_R Q^2, k_F k_R Q_0^2)\,.
\label{3l}
\ee

The functions $\hat C_{NS},\hat H^{NS}$  are to be obtained from $C_{NS}, H^{NS}$ by modifying the RHS of Eqs.~(\ref{1.cf},~\ref{hnns}) as follows:\\
in Eq.~(\ref{1.cf})
\bea
\ar(Q^2) &\to& \ar(k_F Q^2)\,,
\label{2.cc0} \\
B^{NLO}_{NS}(n) &\to& B^{NLO}_{NS}(n) + \frac{1}{2}\gamma_{NS}^{(0)}(n) \ln{k_F}\,, \label{bnlo} \\
B^{NNLO}_{NS}(n) &\to& B^{NNLO}_{NS}(n) + \frac{1}{2}\gamma_{NS}^{(1)}(n) \ln{k_F}
+\left(\frac{1}{2}\gamma_{NS}^{(0)} + \beta_0\right)B_{NS}^{NLO}\ln{k_F}  \nonumber \\
&+&\frac{1}{8}\gamma_{NS}^{(0)}\left(\gamma_{NS}^{(0)} + 2\beta_0\right)\ln^2{k_F}\,,
\label{coeffun}
\eea
and in Eq.~(\ref{hnns})
%(or in Eq.~(\ref{hnnsA}))
\bea
\ar(Q^2) &\to& \ar(k_F k_R Q^2),~~ \ar(Q_0^2) \, \to \, \ar(k_F k_R Q_0^2)
\nonumber \\ \nonumber
Z^{NLO}_{NS}(n) &\to& Z^{NLO}_{NS}(n) + \frac{1}{2}\gamma_{NS}^{(0)}(n)\ln{k_R} \\
Z^{NNLO}_{NS}(n) &\to& Z^{NNLO}_{NS}(n) + \frac{1}{2}\gamma_{NS}^{(1)}(n)\ln{k_R}
+ \frac{1}{2}\gamma_{NS}^{(0)}(n)Z_{NS}^{NLO}\ln{k_R} \nonumber \\
&+&\frac{1}{8}\gamma_{NS}^{(0)}\left(\gamma_{NS}^{(0)} + 2\beta_0\right)\ln^2{k_R}\,.\nonumber
%%~~~~~~~~~~ ~~~~~~~~~~~~ \mbox{ in Eq.(\ref{3.21})}
%%~~~~~~~~~~ ~~~~~~~~~~~\,\, \mbox{ in Eq.(\ref{3.3})}
\eea
Actually,while calculating the coefficient functions $B_{NS}$ the renormalization scale dependence was also taken into account by inserting the term $\beta_0\ln{k_R}\cdot \mathrm{RHS}\,\mathrm{of}\,\mathrm{Eq}.\,(\ref{bnlo})$ into the expression given in Eq.~(\ref{coeffun}) and appropriately modifying Eq.~(\ref{2.cc0}).
In this latter case the above expressions given in Eqs.~(\ref{2.cc0}),~(\ref{coeffun})
are replaced by the following ones:
\bea
\ar(Q^2) &\to& \ar(k_F  k_R Q^2),~~ ~~ \ar(Q_0^2) \, \to \, \ar(k_F k_R Q_0^2)
\,, \nonumber\\ %\label{2.cc0N}
%B^{NLO}_{NS}(n) &\to& B^{NLO}_{NS}(n) + \frac{1}{2}\gamma_{NS}^{(0)}(n)\ln{k_F},\nonumber\\ %\label{bnloN}
B^{NNLO}_{NS}(n) &\to& B^{NNLO}_{NS}(n) + \frac{1}{2}\gamma_{NS}^{(1)}(n)\ln{k_F}+\left(\frac{1}{2}\gamma_{NS}^{(0)} + \beta_0\right)B_{NS}^{NLO}\ln{k_F} \nonumber \\
&+&\frac{1}{8}\gamma_{NS}^{(0)}\left(\gamma_{NS}^{(0)}\!+\!2\beta_0\right)\!\ln^2{k_F}
+\left( B^{NLO}_{NS}(n)\!+\!\frac{1}{2}\gamma_{NS}^{(0)}(n)
\ln{k_F}\right)\beta_0\ln{k_R} \,. \nonumber %\label{coeffunN}
\eea

%%%%%%%%%%%%%%%%%%%%%%%
\subsection{Heavy quark thresholds}

Let's now turn to the problem of threshold crossing. We stick to the so-called variable-flavor-number scheme, in which any heavy quark of the flavor $f$ is considered to be massless and included in the QCD evolution at $Q^2_f$, i.e. $Q^2=Q^2_f$ is the
{\sl threshold point}. The point to cross is taken, following~\cite{CoTu,Shirkov97}, to happen at $Q^2_f=m^2_f$.~\footnote{To be precise, we should have used $m^2_f(Q^2_f)$. However, $m^2_f$ rather weakly depends on $Q^2_f$ in the vicinity of $Q^2_f=m^2_f$. Thus, hereinafter we adopt $m^2_f(m^2_f)=m^2_f$. The values of the heavy quark masses were taken to be those given by the Particle Data Group 2008~\cite{pdg2008}.} % or in some cases at $Q^2_f=4m^2_f$ %just to be able to perform (at least) qualitative comparison with respective results obtained at the NLO level
%(as in our previous work~\cite{KK2001})
A study into other choices for threshold crossing
%another standard choice $Q^2_f=4m^2_f$
and also nowadays popular schemes such as a fixed-flavor one, a general-mass variable-flavor-number one and others (see recent paper~\cite{Alekhin2009} and discussions therein) is deferred to a next paper, with a complete (singlet and nonsinglet) analysis carried out.

Formally, $Q^2$ evolution does not depend on the specific values of $Q^2_0$:
a change in the initial condition from $Q^2_{0,1}$ to $Q^2_{0,2}$ leads only to a change
in the normalization from $M^{NS}_n(Q^2_{0,1})$ to $M^{NS}_n(Q^2_{0,2})$. This
property is well reproduced in our analyses (see discussions at the beginning of Sect.~4). However, the expression for $Q^2$ evolution does depend on the specific values of $Q^2_0$.

{\bf 1.} Let $Q^2_0$ be placed in-between the thresholds of $f$ and $f+1$ flavors, i.e. $Q^2_0=Q^2_0(f)$. Then, the standard evolution
\bea
\frac{M^{NS}_n(f,Q^2)}{M^{NS}_n(f,Q^2_0(f))} &=&
\frac{C^{twist2}_{NS}(n,f,a^{f}_s(Q^2))}{
C^{twist2}_{NS}(n,f,a^{f}_s(Q^2_{0}))}\times
\frac{{\bf f}_{NS}(n,f,Q^2)}{{\bf f}_{NS}(n,f,Q^2_{0}(f))},
%\nonumber
\label{h.1}
\eea
is correct for $Q^2$ values between the thresholds of $f$ and $f+1$ flavors.
Hereafter $a^{f+1}_s(Q^2)$ and $a^{f}_s(Q^2)$ denote the coupling constants above
and below the threshold $Q^2=Q^2_{f+1}$.

As it is well-known in the nonsinglet case the coefficient functions beginning at NNLO level, and anomalous dimensions starting already with NLO, do depend on the number of active quarks. Moreover, starting with NNLO the coupling constant itself is not smooth at $Q^2_f=m^2_f$ (see~\cite{RoSa,Chetyrkin}). Therefore, we have to deal with the modified equations for the latter, which for some heavy quark %~\footnote{A {\bf b}-quark threshold is treated analogously.}
threshold crossing at $Q^2_{f+1}$ are found to be of two options:

{\bf 2.} Consider $Q^2$ evolution above the threshold $Q^2=Q^2_{f+1}$.
Starting from $Q^2=Q^2_{f+1}$, it has the above form (\ref{h.1}) with the replacements
$f \to f+1$ and $Q^2_0(f) \to Q^2_{f+1}$, that is
\bea
\frac{M^{NS}_n(f+1,Q^2)}{M^{NS}_n(f+1,Q^2_{f+1})} &=&
\frac{C^{twist2}_{NS}(n,f+1,a^{f+1}_s(Q^2))}{
C^{twist2}_{NS}(n,f+1,a^{f+1}_s(Q^2_{f+1}))}\times
\frac{{\bf f}_{NS}(n,f+1,Q^2)}{{\bf f}_{NS}(n,f+1,Q^2_{f+1})}\,.
\label{h.1a}
%\nonumber
\eea

The quantity $M^{NS}_n(f,Q^2)$ is observable it should be continuous at
the threshold:
\bea
M^{NS}_n(f+1,Q^2_{f+1}) = M^{NS}_n(f,Q^2_{f+1})\,.
\label{h.2}
\eea
Therefore, the evolution above the threshold $Q^2=Q^2_{f+1}$, in the case of the starting point $Q^2_0$ located below the latter, is found to be of the following form:
\bea
\frac{M^{NS}_n(f+1,Q^2)}{M^{NS}_n(f,Q^2_0(f))}
&=&\frac{C^{twist2}_{NS}(n,f+1,a^{f+1}_s(Q^2))}{
C^{twist2}_{NS}(n,f+1,a^{f+1}_s(Q^2_{f+1}))}
\times \frac{{\bf f}_{NS}(n,f+1,Q^2)}{{\bf f}_{NS}(n,f+1,Q^2_{f+1})}
 \nonumber \\
&\times &\frac{C^{twist2}_{NS}(n,f,a^f_s(Q^2_{f+1}))}{
C^{twist2}_{NS}(n,f,a^f_s(Q^2_0))} \times
\frac{{\bf f}_{NS}(n,f,Q^2_{f+1})}{{\bf f}_{NS}(n,f,Q^2_0(f))}\,.
\label{h.1b}
\eea

{\bf 3.} Below the threshold $Q^2=Q^2_{f}$, we should start from $Q^2=Q^2_{f}$ and use
the expressions given in Eqs.~(\ref{h.1a}) and~(\ref{h.1b}) with the replacements
$f+1 \to f-1$ and $Q^2_{f+1} \to Q^2_{f}$ carried out, i.e.
\bea
\frac{M^{NS}_n(f-1,Q^2)}{M^{NS}_n(f-1,Q^2_{f})} &=&
\frac{C^{twist2}_{NS}(n,f-1,a^{f-1}_s(Q^2))}{
C^{twist2}_{NS}(n,f-1,a^{f-1}_s(Q^2_{f}))}\times
\frac{{\bf f}_{NS}(n,f-1,Q^2)}{{\bf f}_{NS}(n,f-1,Q^2_{f})}. \nonumber
\eea
and
%using (\ref{h.2}) with $f \to f-1$
\bea
\frac{M^{NS}_n(f-1,Q^2)}{M^{NS}_n(f,Q^2_0(f))}
&=&\frac{C^{twist2}_{NS}(n,f-1,a^{f-1}_s(Q^2))}{
C^{twist2}_{NS}(n,f-1,a^{f-1}_s(Q^2_{f}))}
\times \frac{{\bf f}_{NS}(n,f-1,Q^2)}{{\bf f}_{NS}(n,f-1,Q^2_{f})}
\nonumber \\ \nonumber
&\times &\frac{C^{twist2}_{NS}(n,f,a^f_s(Q^2_{f}))}{
C^{twist2}_{NS}(n,f,a^f_s(Q^2_0))} \times
\frac{{\bf f}_{NS}(n,f,Q^2_{f})}{{\bf f}_{NS}(n,f,Q^2_0(f))}\,.
\eea

{\bf 4.} By analogy, in the case of two thresholds situated at $Q^2=Q^2_{f+2}$
and $Q^2=Q^2_{f+1}$ and the initial point of the evolution $Q^2_0$ being
below the threshold $Q^2=Q^2_{f+1}$, the expression is prescribed to be
\bea
\frac{M^{NS}_n(f+2,Q^2)}{M^{NS}_n(f,Q^2_0(f))}
&=&\frac{C^{twist2}_{NS}(n,f+2,a^{f+2}_s(Q^2))}{
C^{twist2}_{NS}(n,f+2,a^{f+2}_s(Q^2_{f+2}))}
\times \frac{{\bf f}_{NS}(n,f+2,Q^2)}{{\bf f}_{NS}(n,f+2,Q^2_{f+2})}
 \\ \nonumber
&\times & \frac{C^{twist2}_{NS}(n,f+1,a^{f+1}_s(Q^2_{f+2}))}{
C^{twist2}_{NS}(n,f+1,a^{f+1}_s(Q^2_{f+1}))}
\times \frac{{\bf f}_{NS}(n,f+1,Q^2_{f+2})}{{\bf f}_{NS}(n,f+1,Q^2_{f+1})}
 \\ \nonumber
&\times &\frac{C^{twist2}_{NS}(n,f,a^f_s(Q^2_{f+1}))}{
C^{twist2}_{NS}(n,f,a^f_s(Q^2_0))} \times
\frac{{\bf f}_{NS}(n,f,Q^2_{f+1})}{{\bf f}_{NS}(n,f,Q^2_0(f))}\,.
\eea

In the case of two thresholds situated at $Q^2=Q^2_{f-1}$
and $Q^2=Q^2_{f}$ and the initial point of the evolution $Q^2_0$ being
%below
above the threshold $Q^2=Q^2_{f-1}$, the rule to follow looks
\bea
\frac{M^{NS}_n(f-2,Q^2)}{M^{NS}_n(f,Q^2_0(f))}
&=&\frac{C^{twist2}_{NS}(n,f-2,a^{f-2}_s(Q^2))}{
C^{twist2}_{NS}(n,f-2,a^{f-2}_s(Q^2_{f-1}))}
\times \frac{{\bf f}_{NS}(n,f-2,Q^2)}{{\bf f}_{NS}(n,f-2,Q^2_{f-1})}
 \\ \nonumber
&\times & \frac{C^{twist2}_{NS}(n,f-1,a^{f-1}_s(Q^2_{f-1}))}{
C^{twist2}_{NS}(n,f-1,a^{f-1}_s(Q^2_{f}))}
\times \frac{{\bf f}_{NS}(n,f-1,Q^2_{f-1})}{{\bf f}_{NS}(n,f-1,Q^2_{f})}
 \\ \nonumber
&\times &\frac{C^{twist2}_{NS}(n,f,a^f_s(Q^2_{f}))}{
C^{twist2}_{NS}(n,f,a^f_s(Q^2_0))} \times
\frac{{\bf f}_{NS}(n,f,Q^2_{f})}{{\bf f}_{NS}(n,f,Q^2_0(f))}\,.
\eea

An extension to the case with any number of thresholds is trivial.

The threshold crossing effect on the coupling constants
is implemented according to the following equations~\cite{RoSa,Chetyrkin}:
\bea
\frac{a^f_s(Q^2_{f+1})}{a^{f+1}_s(Q^2_{f+1})}\!\!\!&=&\!\!\!
1- \frac{2}{3} \ell_{f+1} a^{f+1}_s(Q^2_{f+1}) +
\frac{4}{9} {\left(a^{f+1}_s(Q^2_{f+1})\right)}^2 \, \left[
\ell_{f+1}^2 - \frac{57}{2} \ell_{f+1} + \frac{11}{2} \right], \\
\frac{a^{f+1}_s(Q^2_{f+1})}{a^{f}_s(Q^2_{f+1})}\!\!\!&=&\!\!\!
1+ \frac{2}{3} \ell_{f+1} a^{f}_s(Q^2_{f+1}) +
\frac{4}{9} {\left(a^{f}_s(Q^2_{f+1})\right)}^2 \, \left[
\ell_{f+1}^2 + \frac{57}{2} \ell_{f+1} - \frac{11}{2} \right],
\eea
where
%each term on the RHS is related to the corresponding order of perturbative
%expansion, the quantities found there are
$
\ell_{f+1}=\ln(Q^2_{f+1}/m^2_{f+1})
%\mu_h^2)
%\,,\,\, c_2=\frac{11}{72}\,,\,\,
%c_3=-\frac{82043}{27648}\zeta(3) +
%\frac{575263}{124416}-\frac{2633}{31104}f\,,
%%+\frac{564731}{124416}-\frac{2633}{31104}n_l\,,\,\, n_l=f-1\,,
$.
%and
%%$a', a$ are the coupling constants corresponding to $f-1, f$ quark flavors,
%%respectively. The renormalization group invariant $\MSbar$ mass
%%is taken to be
%$\mu_h \equiv m_h(\mu_h)\approx m_h$.

%%%%%%%%%%%%%%%%%%%%%%%
\subsection{Other aspects of the fits }

Analysis's conditions concerning PDF normalization, target mass (TMC) and higher twist corrections (HTCs), as well as nuclear effects remain essentially the same as in our previous work~\cite{KK2001} so we refer to it for further details, though quoting some salient points.

The moments ${\bf f}_i(n,Q^2)$ at some $Q^2_0$ is a theoretical input to the analysis
which is  fixed as follows.
In the fits of data with the cut $x\geq 0.25$ imposed only the nonsinglet parton
density is worked with and the following patametrization at the normalization point is used (see, for example,~\cite{PKK,KPS}):
\bea
{\bf f}_{NS}(n,Q_0^2) &=& \int_0^1 dx x^{n-2} \tilde{\bf f}_{NS}(x,Q_0^2),
\nonumber \\
\tilde{\bf f}_{NS}(x,Q_0^2) &=& A_{NS}(Q_0^2)
(1-x)^{b_{NS}(Q_0^2)}
(1+d_{NS}(Q_0^2)x)\,,\label{4}
\eea
where $A_{NS}(Q_0^2)$, $b_{NS}(Q_0^2)$ and $d_{NS}(Q_0^2)$ are some
coefficients~\footnote{Here we do not consider the term $\sim x^{a_{NS}(Q_0^2)}$ in the normalization of $\tilde {\bf f}_{NS}(x,Q_0^2)$, because of the cut $x\geq 0.25$.
The correct small-$x$ asymptotics of the nonsinglet distributions is given by Eq.~(29) in~\cite{KK2001} from the corresponding parameters of the valence quark distributions (see Eq.~(26) in~\cite{KK2001}) analyzed with allowance for the complete singlet and nonsinglet evolutions.}.

The distributions of light $\bf{u}$ and $\bf{d}$ quarks,
$\tilde {\bf f}_{u}(x,Q_0^2) \equiv {\bf u}(x,Q_0^2)$ and
$\tilde {\bf f}_{d}(x,Q_0^2) \equiv {\bf d}(x,Q_0^2)$, are composed of two
components:
the valence part --- ${\bf u}_v(x,Q_0^2)$ and ${\bf d}_v(x,Q_0^2)$,
and the sea one --- ${\bf u}_s(x,Q_0^2)$ and ${\bf d}_s(x,Q_0^2)$. For the remaining
quark and antiquark densities only the sea parts are retained. Moreover, following~\cite{Buras,CDR} an equality of all sea parts are assumed with their sum denoted by $S(x,Q_0^2)$.

%===============================================================================
\section{ A fitting procedure }
\label{sec3}

To cut short this follows along the lines described in the previous paper~\cite{KK2001}.
Let's here just recall salient points of the so-called polynomial expansion method.
The latter was first proposed in~\cite{Ynd} and further developed in~\cite{gon}. In these papers the method was based on the Bernstein polynomials and subsequently used to analyze data at NLO~\cite{KaKoYaF,KaKo} and NNLO level~\cite{SaYnd,KPS1}.
The Jacobi polynomials for that purpose were first proposed and then subsequently developed in~\cite{Barker,Kri,Kri1} and used in~\cite{PKK}-\cite{KPS1},~\cite{NNLOBlumlein,Vovk}.

With the QCD expressions for the Mellin moments $M_n^{k}(Q^2)$ analytically calculated according to
the formul\ae\, given above the SF $F_2^k(x,Q^2)$ is reconstructed by using the Jacobi polynomial expansion method:
$$
F_{2}^k(x,Q^2)=x^a(1-x)^b\sum_{n=0}^{N_{max}}\Theta_n ^{a,b}(x)\sum_{j=0}^{n}c_j^{(n)}(\alpha ,\beta )
M_{j+2}^k (Q^2)\,,
\label{2.1}
$$
where $\Theta_n^{a,b}$ are the Jacobi polynomials, $a,b$ are the parameters fitted, and the superscript $k$ is defined in the text just before Eq.~(\ref{1.a}). A condition
put on the former is the requirement of the error minimization while reconstructing the structure functions.

Since a twist expansion starts to be applicable only above $Q^2 \sim 1$ GeV$^2$
the cut $Q^2 \geq 1$ GeV$^2$ on data is applied throughout.

MINUIT program~\cite{MINUIT} is used to minimize two variables
$$
\chi^2_{SF} = \biggl|\frac{F_2^{exp} - F_2^{th}}{\Delta F_2^{exp}}\biggr|^2\,,
\qquad
\chi^2_{slope} = \biggl|\frac{D^{exp} - D^{th}}{\Delta D^{exp}}\biggr|^2\,,
$$
where $D=d\ln F_2/d\ln\ln Q^2$. Quality of the fits is characterized by
$\chi^2/DOF$ for the structure function $F_2$.
%The dependence $F_2 \sim {\left(Q^2\right)}^{-\tilde{d}} $, used in the previous NLO analysis carried out in~\cite{KK2001}, was shown to be a fairly good approximation, which can be explained, in part, by the kinematical relations between the variables $x$ and $Q^2$ (see~\cite{KP02} and discussions in~\cite{KK2009}), as opposed to NNLO analysis, that exposed considerable fit quality deterioration for this particular approximate expression for $F_2$.
%Therefore, we decided to perform also the analysis for another SF slope
%$D$ that serve the purpose of checking the properties of fits, for it sizably reduces
%the correlations between $x$ and $Q^2$, thus allowing a more consistent analysis
%to be carried out.
%More specifically, the quantity $\chi^2_{slope}$ indicates that the dependence $F_2 \sim {\left(a_s(Q^2)\right)}^d$ (see Eq.~(\ref{3}), for example) is actually
%close to $F_2 \sim {\left(\ln(Q^2/\Lambda^2)\right)}^{-d} = \exp \left[-d
%\ln\ln(Q^2/\Lambda^2)\right]$, which  can be considered as appropriate one to use.
However, the analysis show that the experimental data for $F_2$
are strongly correlated in $x$ and $Q^2$; therefore, it is desirable to have at one's disposal some additional characteristics which helps assess the fit quality.
From QCD (see subsection 2.2) it follows that the behaviour
$F_2 \sim {\left(a_s(Q^2)\right)}^d$ (see Eq.~(\ref{3}), for example)
with some $d$ values can be taken to be some crude approximation for $Q^2$-dependence
of the structure function. This form is in a sense similar to $F_2 \sim {\left(\ln(Q^2/\Lambda^2)\right)}^{-d}=\exp\left[-d\ln\ln(Q^2/\Lambda^2)\right]$,
which can be considered as a more appropriate one to use.
In other words, the slope $D\sim -d$ is approximately $Q^2$-independent and, therefore,
suffers rather mild correlations between $x$ and $Q^2$.

%For $D^{th}$ we can take $D=d\ln F_2/d\ln\ln Q^2$ itself. 
The quantities $D^{th}$ and $D^{exp}$, corresponding to ``{\sl experimental data}", can be consructed in the following way. Taking several points of the experimental data for $F_2$ with the same values of $x_i$, we can parametrize them in the form $F_2 \sim {\left(\ln(Q^2/\Lambda^2)\right)}^{-d(x_i)}$.
Then, we can consider the derivative of this expression with respect to $\ln\ln(Q^2/\Lambda^2)$, fitting over each subset of $Q^2_j$ with the average value of the latter obtained by summing up with the weigth $1/F(x_i,Q^2_j)$, and then summing over $x_i$. As a result, basically the following expression is used
$$
D = \sum_{i} \, \frac{\ln(\overline Q^2/\Lambda^2)}{F(x_i,\overline Q^2)} \, \frac{dF(x_i,\overline Q^2)}{d\ln(\overline Q^2/\Lambda^2)}.
$$
The importance of inclusion of the new characteristics into analysis
is shown in Table~4. There it is seen that by adding to the fit step-by-step TMC, HTC and systematic errors, its quality gradually increases. Indeed, the standard $\chi^2_{SF}/DOF$ demonstrates the fit quality improvement, i.e., decreasing from $4.85$ to $0.73$, while the additional $\chi^2_{slope}/DOF$ does it even stronger dropping from $55.33$ all the way down to $0.71$.

%Briefly speaking, dependence of the first mentioned variable on
%the cuts imposed on $Q^2$ for various sets of experimental data is studied both
%with and without higher twist corrections taken into account.

%=======================================================================
\section{Results}

Since there are no gluons in the nonsinglet case the analysis is essentially
easier to conduct. Hence the cut on Bjorken variable ($x\geq 0.25$) imposed
where gluon density is believed to be negligible.

We use free normalizations of the data for different experiments.
For a reference set, the most stable deuterium BCDMS data at the value of the beam initial energy $E_0=200$ GeV is used. With the other data sets taken to be a reference one the variation in the results is still negligible.
In the case of the fixed normalization for each and all data sets the fits tend to yield a little bit worse $\chi^2$, just as before.

The starting point of the evolution is taken to be
$Q^2_0$ = 90 GeV$^2$ for BCDMS data as well as for overall data
and $Q^2_0$ = 20 GeV$^2$ --- for the combined SLAC, NMC and BFP data.
%The choice of $Q^2_0$-values  is in good agreement with above conditions (see the previous Section).
These $Q^2_0$ values are close to the average values of $Q^2$ spanning the corresponding data.
To check for $Q^2_0$-independence we use also other $Q^2_0$ values: $Q^2_0$ =
2 GeV$^2$ and $Q^2_0$ = 10 GeV$^2$. We find that a variation of the results,
presented below, is of the order of ${\cal O}(10^{-5})$ for the values of $\asMZ$ and, therefore, to the accuracy we work in can be said to be negligible.

On grounds of previous knowledge the maximal value of the number
of moments to be accounted for is $N_{max} =8$~\cite{Kri,Kri1} (though we
check $N_{max}$ dependence like in the NLO analysis) and
the cut $0.25 \leq x \leq 0.8$ is imposed everywhere.

\subsection { BCDMS data with carbon, hydrogen and deuterium targets}

Analysis commences on with the most precise experimental data~\cite{BCDMS1,BCDMS2,BCDMS3}
obtained  by BCDMS muon scattering experiment for large $Q^2$ values.
A complete set of data includes 607 points for the lower cut $x \geq 0.25$.
As was pointed out earlier the starting point of QCD evolution is $Q^2_0=90$ GeV$^2$.
%To proceed with comparison to NLO results
The heavy quark thresholds are %firstly
taken to be %at $Q^2_f=4m^2_f$ (Table~2) and then
at $Q^2_f=m^2_f$ (Table~2).
An original analysis carried out by BCDMS collaboration (see
also~\cite{ViMi}) gave (back then) comparatively small values for the strong coupling constant; for example, $\alpha_s(M^2_Z)=0.113$ at NLO was quoted in the latter reference.

Just like in our previous work~\cite{KK2001} an issue with the data systematic
errors still remains. Let's impose cuts on the kinematic variable $Y=(E_0-E)/E_0$,
where $E_0$ and $E$ are lepton's initial and final energies, respectively~\cite{Kri2}.
Upon excluding a set of data with large systematic errors
considerably higher values of $\alpha_s(M^2_Z)$ are obtained and rather mild
dependence of its values on the choice of $Y$ cut is observed.
For more details we refer to~\cite{KK2001}.

Impact of experimental systematic errors on the results of QCD analysis
as a function of $Y_{cut3}$, $Y_{cut4}$ and $Y_{cut5}$ imposed on data is studied.
The following $y$ cuts depending on the limits put on $x$ are applied:
\bea
& &y \geq 0.14 \,~~~\mbox{ for  }~~~ 0.3 < x \leq 0.4 \nonumber \\
& &y \geq 0.16 \,~~~\mbox{ for  }~~~ 0.4 < x \leq 0.5 \nonumber \\
& &y \geq Y_{cut3} ~~~\mbox{ for  }~~~ 0.5 < x \leq 0.6 \nonumber \\
& &y \geq Y_{cut4} ~~~\mbox{ for  }~~~ 0.6 < x \leq 0.7 \nonumber \\
& &y \geq Y_{cut5} ~~~\mbox{ for  }~~~ 0.7 < x \leq 0.8 \nonumber
%\label{cut}
\eea
Several cases for the three last conditions, with the cut $0.5 < x \leq 0.8$  imposed on the Bjorken variable, are presented in Table~1.

\vspace{0.5cm}
{\bf Table 1.} {\sl A set of $Y_{cut3}$, $Y_{cut4}$ and $Y_{cut5}$ values used in the analysis}
\begin{center}
\begin{tabular}{|c|c|c|c|c|c|c|c|}
\hline
& & & & & & & \\
$N_{Y_{cut}}$ & 0 & 1 & 2 & 3 & 4 & 5 & 6 \\
& & & & & & & \\
\hline \hline
$Y_{cut3}$ & 0 & 0.14 & 0.16 & 0.16 & 0.18 & 0.22 & 0.23 \\
%\hline
$Y_{cut4}$ & 0 & 0.16 & 0.18 & 0.20 & 0.20 & 0.23 & 0.24 \\
$Y_{cut5}$ & 0 & 0.20 & 0.20 & 0.22 & 0.22 & 0.24 & 0.25 \\
\hline
\end{tabular}
\end{center}
\vspace{0.5cm}

The systematic errors for BCDMS data are given~\cite{BCDMS1,BCDMS2,BCDMS3}
as multiplicative factors to be applied to $F_2(x,Q^2)$: $f_r, f_b, f_s, f_d$
and $f_h$ are the uncertainties caused by the spectrometer resolution, beam momentum,
calibration, spectrometer magnetic field calibration, detector inefficiencies
and the energy normalization, respectively.

Each experimental point of the original data set was multiplied
by a factor characterizing the type of uncertainties under consideration
and then the data set modified that way was once again fitted along the lines
of the procedure given in the previous section.
The factors $f_r, f_b, f_s, f_d, f_h$ were read off from~\cite{BCDMS1,BCDMS2,BCDMS3}.
Absolute differences between the $\alpha_s$ values for both original and modified data sets
are shown in Table~2 in the column for a total systematic error estimated in quadrature.
There as well given are the number of experimental points and $\alpha_s$ value
for the initial data set.

\vskip0.5cm
{\bf Table 2.} $\asMZ$ {\sl values for various sets of $Y$ cuts imposed on the data}
\begin{center}
\begin{tabular}{|l|c|c|c|c|c|}
\hline
& & & & & \\
$N_{Y_{cut}}$  & number & $\chi^2(F_2)/DOF$  & $\as(90~\mbox{GeV}^2)$
&total &  $\as(M_Z^2)$  \\
&of points &  &  $\pm$ stat. error  & syst. error & $\pm$ stat. error \\
\hline \hline
0 & 607 & 1.06 & 0.1523 $\pm$ 0.0025 & 0.0136 & 0.1056 $\pm$ 0.0012 \\
1 & 511 & 0.96 & 0.1671 $\pm$ 0.0033 & 0.0103 & 0.1123 $\pm$ 0.0014 \\
2 & 502 & 0.96 & 0.1680 $\pm$ 0.0034 & 0.0097 & 0.1127 $\pm$ 0.0015 \\
3 & 495 & 0.95 & 0.1685 $\pm$ 0.0034 & 0.0094 & 0.1129 $\pm$ 0.0015 \\
4 & 489 & 0.95 & 0.1701 $\pm$ 0.0035 & 0.0091 & 0.1136 $\pm$ 0.0015 \\
5 & 458 & 0.94 & 0.1719 $\pm$ 0.0037 & 0.0078 & 0.1144 $\pm$ 0.0016 \\
6 & 452 & 0.93 & 0.1729 $\pm$ 0.0037 & 0.0075 & 0.1148 $\pm$ 0.0016 \\
\hline
\end{tabular}
\end{center}
% for Mf=2mf
%0 & 607 & 0.98 & 0.1454 $\pm$ 0.0024 & 0.0129 & 0.1022 $\pm$ 0.0011 \\
%1 & 511 & 0.96 & 0.1496 $\pm$ 0.0029 & 0.0090 & 0.1043 $\pm$ 0.0013 \\
%2 & 502 & 0.96 & 0.1503 $\pm$ 0.0030 & 0.0085 & 0.1046 $\pm$ 0.0013 \\
%3 & 495 & 0.95 & 0.1508 $\pm$ 0.0030 & 0.0081 & 0.1048 $\pm$ 0.0013 \\
%4 & 489 & 0.95 & 0.1520 $\pm$ 0.0031 & 0.0076 & 0.1054 $\pm$ 0.0013 \\
%5 & 458 & 0.93 & 0.1530 $\pm$ 0.0033 & 0.0068 & 0.1059 $\pm$ 0.0014 \\
%6 & 452 & 0.93 & 0.1539 $\pm$ 0.0034 & 0.0065 & 0.1063 $\pm$ 0.0014 \\

For illustrative purposes let's depict these numbers (to be precise, for~$\alpha_s(M_Z^2)$) in Fig.~2 with NLO results (evaluated in this work) included for comparison.
It is seen that the value of the coupling is less than in NLO throughout
as was generally expected. Also note bigger systematic errors with respect to the previous analysis which can presumably be ascribed to the scheme of threshold crossing used.
%\vspace{0.5cm}
%\newpage
\begin{figure}[!htb] %\label{fig-1}
\unitlength=1mm
\vskip -1.5cm
\begin{picture}(0,100)
  \put(0,-5){%
   \psfig{file=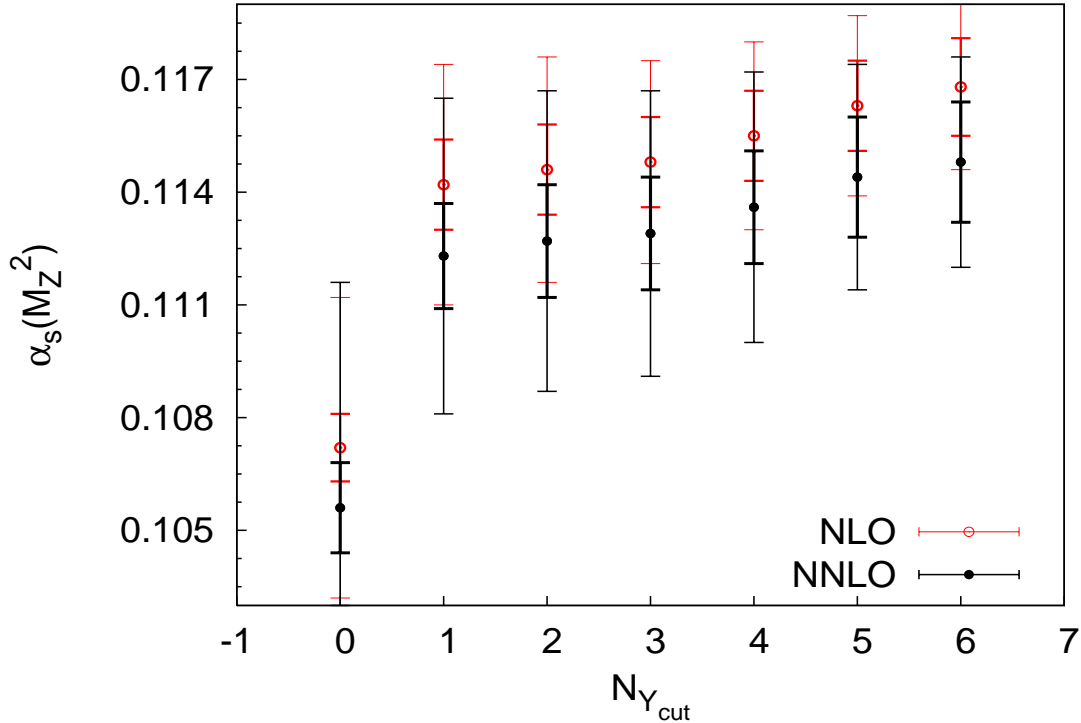,width=150mm,height=100mm}
}
\end{picture}
\vskip 0.2cm
\caption{\it{
Effect of systematic errors on the value of the coupling constant for different $Y_{cut}$ values in the fits based on the nonsinglet evolution.
Data analyzed are BCDMS $C^{12}, H_2, D_2$ sets with the cuts $x_{min}=0.25$ and those from Table~1 imposed. The starting point for evolution is taken to be $Q_0^2=90$ GeV$^2$. Thresholds of $\bf c$ and $\bf b$ quarks are chosen to be $Q^2_{\bf c}=1.61$ GeV$^2$ and $Q^2_{\bf b}=17.64$ GeV$^2$, respectively. The inner (outer) error bars show statistical (systematic) errors}.}
\end{figure}
%\vspace{0.5cm}

From the figure one can observe that similar to the analysis done
at NLO level the values of $\alpha_s$ are stable and statistically
consistent throughout an entire set of $Y$ cuts imposed on the
data, though there is a slight trend in the central values of the
coupling to increase towards higher $N_{Y_{cut}}$.
As in the earlier analysis the case $N_{Y_{cut}}=6$ is once again most
attractive for reducing a total systematic error in $\alpha_s$ by nearly half as much.
At the same time, increase of the statistical error by 50\% is observed just like in NLO case.
%The figures for errors themselves are reduced by more than a half as compared to those of the NLO analysis.

Upon the cuts imposed (in what follows we use the set $N_{Y_{cut}}=6$), only 452 points left available. Fitting them according to the procedure outlined above the following results are obtained:
\bea
\as(90~\mbox{GeV}^2) &=& 0.1729 \pm 0.0037 ~\mbox{(stat)}
\pm 0.0075 ~\mbox{(syst)} \pm 0.0016~\mbox{(norm)}
\nonumber \\
& & \label{bd1.a} \\
\as(M_Z^2) &=& 0.1148 \pm 0.0016~\mbox{(stat)}
\pm 0.0030~\mbox{(syst)} \pm 0.0007~\mbox{(norm)},
\nonumber
\eea
where onwards an abbreviate ``norm'' denotes the experimental data normalization
error which
comes from the difference of the fits with free and fixed normalizations of BCDMS data subsets~\cite{BCDMS1,BCDMS2,BCDMS3} with different values of the beam energy.
Therefore, for the fits of BCDMS data in the case of NS evolution and
under a condition of minimizing systematic errors the following results are obtained:
\be
\as(M_Z^2) ~=~ 0.1148\pm 0.0035~\mbox{(total exp. error)}\,,
\label{bd1.a1}
\ee
where an estimate for the total experimental error comes from the
statistical, systematic and normalization errors taken in quadrature.
Note that this figure is higher than that given by BCDMS itself and quoted at the beginning of this subsection.

Similar to the case of NLO analysis let's scrutinize the dependence of results on a maximal number of polynomials $N_{max}$ used in fits.
A full set of data is comprised of 452 points with the $Q^2$-evolution starting from
$Q^2_0$=90 GeV$^2$. As it can be seen from Table~3 similar sort of stability of the results still holds in good agreement with~\cite{Kri}.

As it is seen from Table~3 beginning with $N_{max}=4$ the resulting values obtained are rather stable; therefore, an average value of the coupling constant can be calculated and is found to be $\as(M_Z^2)=0.1145$. Average deflection is estimated to be $0.0003$ and can be considered a method error.
%A few words on a slope behavior are in order. Recall that at NLO level the proposed dependence $F_2\sim Q^{2(\gamma)}$ was a good approximation and check for the property of the fits. This cannot be said given results obtained in the present paper. On the other hand, the quantity $\chi^2_{slope1}$ seems to indicate that the dependence $F_2\sim [\ln Q^2]^\gamma$ is to be considered a more appropriate one to use.
%Thus, the role of NNLO corrections in shaping the form of the DIS structure function dependence on the variables $x,Q^2$ is getting even further obscure.

\subsection{ SLAC, NMC (hydrogen and deuterium), and BFP (iron) data sets}

NS evolution analysis is continued with fitting the experimental data
obtained by SLAC, NMC and BFP collaborations~\cite{SLAC1,SLAC2,NMC,BFP}.
A full set of data upon imposing a cut $x \geq 0.25$ consists of 345 points:
238 SLAC points, 66 NMC points and 41 those of BFP.
The starting point of the QCD evolution is $Q^2_0=20$ GeV$^2$ and the $Q^2$-cut imposed
is $Q^2>1$ GeV$^2$. For SLAC and NMC data the statistical and systematic errors are combined in quadrature.

\vspace{0.5cm}
{\bf Table 3.} $\asMZ$ {\sl for various} $N_{max}$ {\sl values}
\begin{center}
\begin{tabular}{|c|c|c|c|c|}
\hline
& & & & \\
$N_{max}$ & $\chi^2(F_2)/DOF$ &
%$ \chi^2_{slope}$ &
$ \chi^2_{slope} $
& $\as(90~\mbox{GeV}^2)$ & $\asMZ$ \\
&
%& for $6$ points
& for $6$ points & $\pm$ 0.0037 &  $\pm$ 0.0012 \\
\hline \hline
3 & 1.06
%& 7.7
& 4.3 & 0.1691 & 0.1132 \\
%\hline
4 & 0.96
%& 14.8
& 5.5 & 0.1708 & 0.1139  \\
%\hline
5 & 0.96 &
%18.0&
6.6 & 0.1702 & 0.1137 \\
%\hline
6 & 0.93 &
%7.0 &
5.6 & 0.1741 & 0.1154 \\
%\hline
7 & 0.93 &
%9.4 &
4.6 & 0.1732 & 0.1150 \\
%\hline
8 & 0.93 &
%10.7&
5.0 & 0.1729 & 0.1148 \\
%\hline
9 & 0.93 &
%11.2&
5.4 & 0.1727 & 0.1148  \\
%\hline
10 & 1.05 &
%11.5 &
7.0 & 0.1726 & 0.1147 \\
%\hline
11 & 1.06 &
%10.9 &
5.4 & 0.1716 & 0.1143 \\
%\hline
12 & 1.02 &
%10.6 &
5.7 & 0.1717 & 0.1143 \\
%\hline
13 & 1.10 &
%11.4 &
5.7 & 0.1715 & 0.1142 \\
%3 & 1.06 & 7.8 & 6.5 & 0.1609 & 0.1095 \\
%4 & 0.96 & 18.4& 9.2 & 0.1643 & 0.1111  \\
%5 & 0.96 & 21.9& 10.9 & 0.1637 & 0.1108 \\
%\hline
%6 & 0.94 & 7.6 & 7.7 & 0.1652 & 0.1115 \\
%\hline
%7 & 0.93 & 11.1& 7.7 & 0.1654 & 0.1116 \\
%\hline
%8 & 0.94 & 12.1& 7.9 & 0.1653 & 0.1115 \\
%\hline
%9 & 0.93 & 13.4& 8.3 & 0.1652 & 0.1115  \\
%\hline
%10 & 1.06 & 13.6 & 10.2 & 0.1649 & 0.1113 \\
%\hline
%11 & 1.06 & 13.5 & 9.1 & 0.1645 & 0.1111 \\
%\hline
%12 & 1.10 & 12.8 & 8.0 & 0.1645 & 0.1112 \\
%\hline
%13 & 1.10 & 13.4 & 8.3 & 0.1644 & 0.1111 \\
\hline
\end{tabular}
\end{center}
\vspace{0.5cm}

To illustrate importance of $1/Q^2$ corrections the fits of the data are performed in the following way. Firstly, one compares the data with the perturbative QCD part of SF $F_2$, i.e. $F_2^{twist2}$ taken into account.
Then, $1/Q^2$ corrections beginning with target mass ones are added followed by the
account for the twist-four terms.
As it is easy to read off from Table~4 we have unsatisfactory fit
when we work with the leading twist part $F_2^{twist2}$ only.
Agreement with the data appears to be improving upon including into analysis
the target mass corrections. Eventually an allowance for the twist-four corrections
leads to a very good fit of the data.
Also, it is seen that the results are considerably spoilt by the neglect of systematic errors for SLAC and NMC data, just like those obtained in NLO analysis.
%It is also seen that neglect of systematic errors considerably spoil quality of the fit.

\vspace{0.5cm}
{\bf Table 4.} $\asMZ$ and $\chi^2$ {\sl for various fits with/without TMC, HTC, and
systematic errors}
\begin{center}
\begin{tabular}{|l|c|c|c|c|c|c|c|c|}
\hline
& & & & & & & \\
$N$  & TMC & HTC & syst. &$\chi^2(F_2)/DOF$ & $\chi^2_{slope}$
& $\as(20~\mbox{GeV}^2)$ & $\asMZ$ \\
 & & & error &  &for $8$ points  &$\pm$ stat  &   \\
\hline \hline
1 & No & No & Yes & 4.85 & 442.6 & 0.2260 $\pm$ 0.0015 & 0.1197 \\
%\hline
2 & Yes & No & Yes & 2.05& 87.6 & 0.2054 $\pm$ 0.0014 & 0.1139 \\
%\hline
%3 & Yes & Yes & No & 1.50 & 9.5 & 0.2328 $\pm$ 0.0020 & 0.1215 \\
3 & Yes & Yes & No & 1.47 & 14.7 & 0.2183 $\pm$ 0.0031 & 0.1176 \\
%\hline \hline
%4 & Yes & Yes & Yes & 0.78 & 3.6 & 0.2342 $\pm$ 0.0025 & 0.1218 \\
4 & Yes & Yes & Yes & 0.73 & 5.7 & 0.2188 $\pm$ 0.0051 & 0.1177 \\
%1 & No & No & Yes & 4.4 & 806 & 0.2058 $\pm$ 0.0012 & 0.1140 \\
%%\hline
%2 & Yes & No & Yes & 1.6 & 96 & 0.1916 $\pm$ 0.0012 & 0.1097 \\
%%\hline
%3 & Yes & Yes & No & 2.0 &19.6 & 0.1957 $\pm$ 0.0043 & 0.1109 \\
%%\hline \hline
%4 & Yes & Yes & Yes & 0.98 & 3.6 & 0.2039 $\pm$ 0.0055 & 0.1134 \\
\hline
\end{tabular}
\end{center}
\vspace{0.5cm}

%The following values for parameters in the parametrizations of parton distributions (at $Q^2_0=20$ GeV$^2$) are obtained:
%\bea
%A_{NS}^{H_2} &=& 2.00,~~~~~~~~ A_{NS}^{D_2} ~=~ 2.44,~~~~~~~~~ A_{NS}^{Fe} ~=~ 1.91,  \nonumber \\
%b_{NS}^{H_2} &=& 3.95,~~~~~~~~~ b_{NS}^{D_2} ~=~ 4.00,~~~~~~~~~ b_{NS}^{Fe} ~=~ 4.45,  \nonumber \\
%d_{NS}^{H_2} &=& 8.86,~~~~~~~~~ d_{NS}^{D_2} ~=~ 5.26,~~~~~~~~~ d_{NS}^{Fe} ~=~ 7.52\,.
%\label{para}
%\eea
%Here the symbols $H_2$, $D_2$ and $Fe$ denote the parametrizations for hydrogen, deuterium and iron data, respectively.
%Note that the values of the coefficients are close to those obtained in other numerical analyses (see~\cite{KPS,KPS1,Alekhin03} and references therein).
%The values of the parameters $b_{NS}^\ell$ ($\ell=H_2,D_2,Fe$) are in good agreement with the quark counting rules~\cite{schot}, as well as with some theoretical studies~\cite{Gross,VoKoMa}.

To conclude the following results for $\chi^2(F_2)=251$ and $\chi^2_{slope}=5.7$
over $8$ points are obtained:
\bea
\as(20~\mbox{GeV}^2) &=& 0.2188 \pm 0.0051 ~\mbox{(stat)}
\pm 0.0084~\mbox{(syst)} \pm 0.0025~\mbox{(norm)} \nonumber \\ & & \label{slo1}
\\ \nonumber \\
\as(M_Z^2) &=& 0.1177 \pm 0.0014~\mbox{(stat)}
%+ \biggl\{ \begin{array}{l}+ 0.0020 \\ - 0.0034 \end{array}
\pm 0.0035~\mbox{(syst)} \pm 0.0008~\mbox{(norm)}\,. \nonumber
%\label{slo1}
\eea
The last error $\pm 0.0008$ to $\as(M_Z^2)$ comes again from the fits with free and fixed normalizations among different data sets provided by the SLAC, NMC and BFP collaborations.

Thus, by combining errors in quadrature the fits based on the nonsinglet evolution
give for the strong coupling constant:
\bea
\as(M_Z^2) ~=~ 0.1177 \pm 0.0039 ~\mbox{(total exp. error)}\,.
\label{bd1.2}
\eea

Looking at the results obtained so far one observes fairly good agreement within errors given between the values of $\as(M_Z^2)$ derived from the fits of BCDMS data alone and those from the fits of combined SLAC, NMC and BFP data. %~\ref{slo1}-\ref{bd11.1} (see Eqs. (\ref{bd1.a})-(\ref{bd11.a}) and (\ref{slo1})-(\ref{bd11.1})).
Let's now put all the data together and fit them simultaneously.

\subsection { SLAC, BCDMS, NMC and BFP data sets }

Just as above for the BCDMS data the cuts imposed are $x\geq 0.25$ along with $Y_{cut}$ and $N_{Y_{cut}}=6$ (see Table~1). Then a full set of data consists of 797 points.
The starting point of the QCD evolution is once again taken to be $Q^2_0=90$ GeV$^2$.

\vspace{0.5cm}
{\bf Table 5.} $\asMZ$ {\sl and} $\chi^2$ {\sl in the case of the combined analysis}
\begin{center}
\begin{tabular}{|l|c|c|c|c|c|c|}
\hline
& &  & &  &\\
$Q^2_{min}$ & $N$ of & HTC &$\chi^2(F_2)$/DOF &
$\as(90~\mbox{GeV}^2)$ $\pm$ stat & $\asMZ$ \\
& points &  &  & &  \\
\hline \hline
1.0 & 797 &  No & 2.20 & 0.1767 $\pm$ 0.0008 & 0.1164 \\
2.0 & 772 &  No & 1.14 & 0.1760 $\pm$ 0.0007 & 0.1162 \\
3.0 & 745 &  No & 0.97 & 0.1788 $\pm$ 0.0008 & 0.1173 \\
4.0 & 723 &  No & 0.92 & 0.1789 $\pm$ 0.0009 & 0.1174 \\
5.0 & 703 &  No & 0.92 & 0.1793 $\pm$ 0.0010 & 0.1176 \\
6.0 & 677 &  No & 0.92 & 0.1793 $\pm$ 0.0012 & 0.1176 \\
7.0 & 650 &  No & 0.92 & 0.1782 $\pm$ 0.0015 & 0.1171 \\
8.0 & 632 &  No & 0.93 & 0.1773 $\pm$ 0.0018 & 0.1167 \\
9.0 & 613 &  No & 0.93 & 0.1764 $\pm$ 0.0022 & 0.1163 \\
10.0 & 602 & No & 0.92 & 0.1742 $\pm$ 0.0023 & 0.1154 \\
%11.0 & 588 & No & 0.91 & 0.1718 $\pm$ 0.0027 & 0.1144 \\
%12.0 & 574 & No & 0.92 & 0.1717 $\pm$ 0.0029 & 0.1143 \\
%13.0 & 570 & No & 0.92 & 0.1710 $\pm$ 0.0030 & 0.1140 \\
%14.0 & 562 & No & 0.92 & 0.1712 $\pm$ 0.0032 & 0.1141 \\
%15.0 & 550 & No & 0.91 & 0.1715 $\pm$ 0.0033 & 0.1142 \\
\hline \hline
1.0 & 797 & Yes & 0.98 & 0.1772 $\pm$ 0.0027 & 0.1167 \\
%1.0 & 797 &  No & 2.34 & 0.1756 $\pm$ 0.0008 & 0.1160 \\
%2.0 & 772 &  No & 1.16 & 0.1760 $\pm$ 0.0007 & 0.1162 \\
%3.0 & 745 &  No & 0.97 & 0.1791 $\pm$ 0.0009 & 0.1175 \\
%4.0 & 723 &  No & 0.93 & 0.1794 $\pm$ 0.0009 & 0.1176 \\
%5.0 & 703 &  No & 0.92 & 0.1799 $\pm$ 0.0011 & 0.1178 \\
%6.0 & 677 &  No & 0.92 & 0.1800 $\pm$ 0.0013 & 0.1178 \\
%7.0 & 650 &  No & 0.92 & 0.1788 $\pm$ 0.0016 & 0.1173 \\
%8.0 & 632 &  No & 0.93 & 0.1778 $\pm$ 0.0018 & 0.1169 \\
%9.0 & 613 &  No & 0.93 & 0.1767 $\pm$ 0.0022 & 0.1165 \\
%10.0 & 602 & No & 0.92 & 0.1746 $\pm$ 0.0024 & 0.1156 \\
%11.0 & 588 & No & 0.91 & 0.1721 $\pm$ 0.0028 & 0.1145 \\
%12.0 & 574 & No & 0.92 & 0.1719 $\pm$ 0.0030 & 0.1144 \\
%13.0 & 570 & No & 0.92 & 0.1712 $\pm$ 0.0030 & 0.1141 \\
%14.0 & 562 & No & 0.92 & 0.1716 $\pm$ 0.0032 & 0.1142 \\
%15.0 & 550 & No & 0.91 & 0.1717 $\pm$ 0.0034 & 0.1143 \\
%16.0 & 542 & No & 0.92 & 0.1719 $\pm$ 0.0034 & 0.1144 \\
%17.0 & 535 & No & 0.92 & 0.1733 $\pm$ 0.0035 & 0.1150 \\
%18.0 & 532 & No & 0.92 & 0.1739 $\pm$ 0.0035 & 0.1153 \\
%19.0 & 525 & No & 0.92 & 0.1738 $\pm$ 0.0036 & 0.1152 \\
%20.0 & 519 & No & 0.92 & 0.1739 $\pm$ 0.0036 & 0.1153 \\
%\hline \hline
%1.0 & 797 & Yes & 1.00 & 0.1762 $\pm$ 0.0029 & 0.1162 \\
\hline
\end{tabular}
\end{center}
\vspace{0.5cm}

To verify a range of applicability of perturbative QCD
we start with analyzing the data without a contribution of twist-four terms
(which means $F_2 = F_2^{pQCD}$) and perform several fits with the cut
$Q^2 \geq Q^2_{min}$ gradually increased.
From Table~5 it is seen that unlike the previous analysis \cite{KK2001}
quality of the fits starts to appear fairly good already from $Q^2=3$ GeV$^2$ onwards.
Except for the order of the approximation at which the analysis is performed the basic difference between this analysis and that carried out in~\cite{KK2001} is in the value of the thresholds that leads to additional increase in the value of the coupling constant in comparison to the case of cutting BCDMS data out with large systematic errors. Thus, a combination of NNLO approximation and the thresholds taken at $Q_f=m_f$ rather than $Q_f=2m_f$ essentially improves agreement between perturbative QCD and the experimental data.

To proceed with comparison, the twist-four corrections are added and the data with the usual cut $Q^2 \geq 1$ GeV$^2$ is fitted. It is clearly seen that as in
the NLO case here the higher twists do sizably improve the quality of the fit,
with insignificant discrepancy in the values of the coupling constant to be quoted below.
% After this, good quality of the fit is observed.
%Besides $\asMZ$ values obtained by both ways of fitting outlined above are quite close to each other (see Table~5).

\begin{figure}[!htb] %\label{HTC_comp}
\unitlength=1mm
\vskip -1.5cm
\begin{picture}(0,100)
  \put(0,-5){%
   \psfig{file=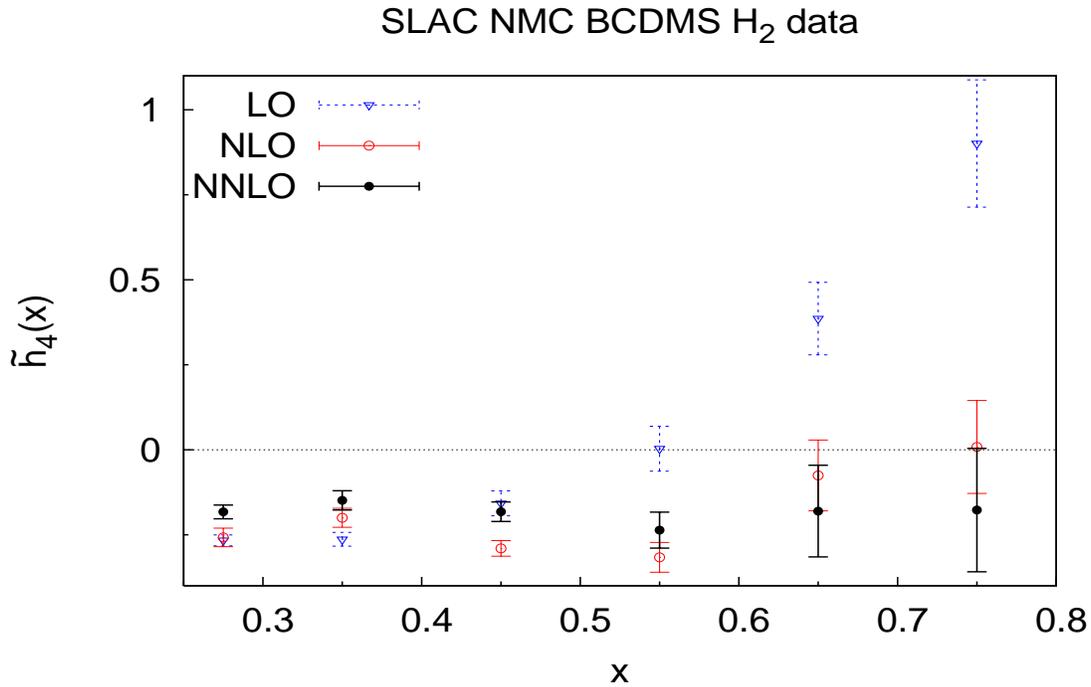,width=150mm,height=95mm}
}
\end{picture}
\vskip 0.2cm
\caption{\it{ Comparison of the HTC parameter $\tilde h_4(x)$ obtained at LO, NLO and NNLO for hydrogen data (the bars stand for statistical errors)}.}
\end{figure}

%\vskip-.5cm
\begin{figure}[!htb] %\label{HTC_comp}
\unitlength=1mm
\vskip -1.5cm
\begin{picture}(0,100)
\put(0,-5){\psfig{file=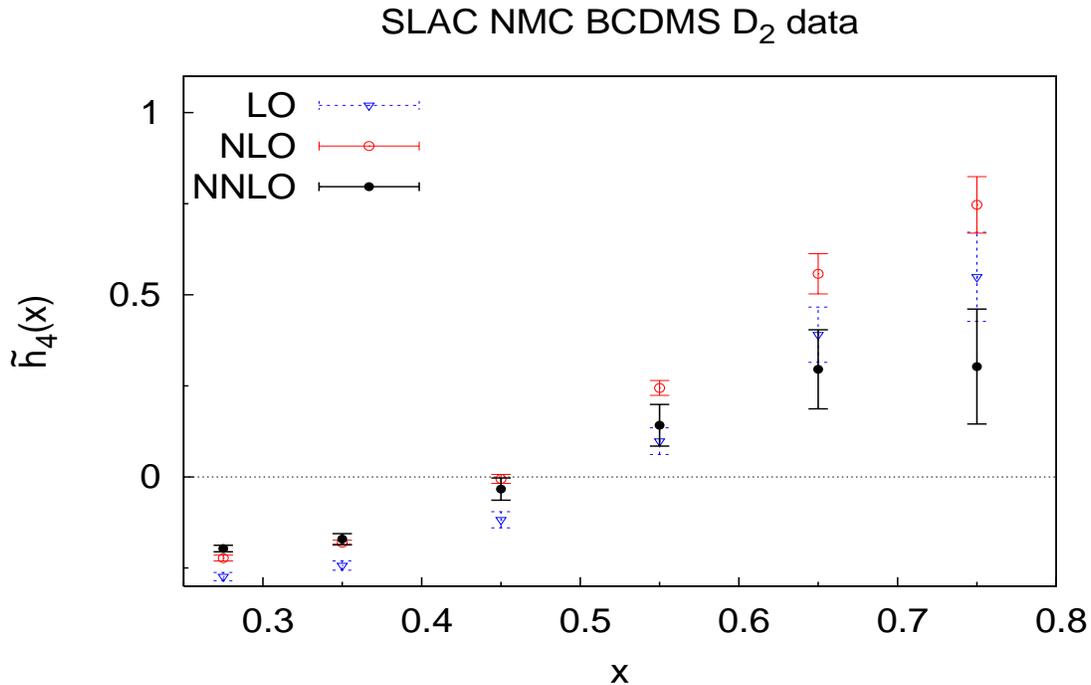,width=150mm,height=95mm}}
\end{picture}
\vskip 0.2cm
\caption{\it{The same as in Fig.~3 for deuterium data}.}
\end{figure}

The following values for the parameters of the parton distribution parametrizations for the case corresponding to the last row of Table~5 are obtained:
\bea
A^{H_2}_{NS} &=& 2.54\pm 0.02\,,~ A^{D_2}_{NS} ~=~ 2.38\pm 0.03\,,~ A^C_{NS} ~=~ 3.29\pm 0.04\,,~ A^{Fe}_{NS} ~=~ 2.35\pm 0.17,  \nonumber \\
b^{H_2}_{NS} &=& 4.16\pm 0.01\,,~ b^{D_2}_{NS} ~=~ 4.22\pm 0.01\,,~ b^C_{NS} ~=~ 4.23\pm 0.03\,,~ b^{Fe}_{NS} ~=~ 4.39\pm 0.21,   \nonumber \\
d^{H_2}_{NS} &=& 6.08\pm 0.17\,,~ d^{D_2}_{NS} ~=~ 3.89\pm 0.12\,,~ d^C_{NS} ~=~ 2.02\pm 0.19\,,~ d^{Fe}_{NS} ~=~ 3.31\pm 1.47\,. \nonumber
\eea
%some of which are on par with those presented in the previous section (see also discussions given there).

The parameter values of the twist-four term are presented in Table~6.
Note that these for $H_2$ and $D_2$ targets are obtained in separate fits by analyzing SLAC, NMC and BCDMS data sets taken together.
It is seen that the values at NLO and NNLO match within errors with an average value being slightly less for the latter.
%The values of this parameter in NLO approximation agrees with those of the analogous analysis carried out in~\cite{ViMi}.

For illustrative purposes we visualize them in Figs.~3, 4 %and~5
where fairly good agreement between higher twist corrections obtained at NLO and NNLO is observed, that is in agreement %not in contradiction
with earlier studies (see, for example,~\cite{Alekhin03}). However, at large $x$ the central values of HTCs are a bit decreased at NNLO level.
%%, especially for hydrogen data.
Moreover, in the case of deuterium data the HT parameter values in LO for large $x$ are less than those obtained in NLO, in contrast to some studies.

\vskip 0.5cm
{\bf Table 6.} {\sl Parameter values of the twist-four term in different orders}
\begin{center}
\scriptsize
\begin{tabular}{|c||c|c||c|c||c|c|}
\hline
 & \multicolumn{2}{c||}{LO $\pm$ stat} & \multicolumn{2}{c||}{NLO $\pm$ stat} & \multicolumn{2}{c|}{NNLO $\pm$ stat}  \\
\cline{2-7}
$x$ & $\tilde h_4(x)$ for $H_2$ & $\tilde h_4(x)$ for $D_2$ & $\tilde h_4(x)$ for $H_2$ & $\tilde h_4(x)$ for $D_2$ & $\tilde h_4(x)$ for $H_2$ & $\tilde h_4(x)$ for $D_2$ \\
% & $\pm$ stat &  $\pm$ stat & $\pm$ stat &  $\pm$ stat & $\pm$ stat &  $\pm$ stat  \\
\hline \hline
0.275&-0.266$\pm$0.016&-0.273$\pm$0.011&-0.258$\pm$0.027&-0.223$\pm$0.008& -0.183 $\pm$ 0.020  & -0.197 $\pm$ 0.009\\
0.35 &-0.263$\pm$0.020&-0.243$\pm$0.013&-0.200$\pm$0.028&-0.181$\pm$0.007& -0.149 $\pm$ 0.028  & -0.171 $\pm$ 0.015 \\
0.45 &-0.157$\pm$0.037&-0.118$\pm$0.022&-0.290$\pm$0.023&-0.005$\pm$0.012& -0.182 $\pm$ 0.029  & -0.033 $\pm$ 0.031 \\
0.55 &0.003 $\pm$0.066&0.098 $\pm$0.037&-0.316$\pm$0.044&0.244 $\pm$0.020& -0.236 $\pm$ 0.052  &  0.142 $\pm$ 0.057 \\
0.65 &0.386 $\pm$0.107&0.390 $\pm$0.076&-0.075$\pm$0.104&0.558 $\pm$0.055& -0.180 $\pm$ 0.135  &  0.295 $\pm$ 0.108 \\
0.75 &0.901 $\pm$0.187&0.549 $\pm$0.122&0.008 $\pm$0.137&0.747 $\pm$0.078& -0.177 $\pm$ 0.182  &  0.303 $\pm$ 0.158 \\
%0.275& -0.163 $\pm$ 0.006 & -0.124 $\pm$ 0.010 \\
%0.35 & -0.133 $\pm$ 0.006 & -0.024 $\pm$ 0.016 \\
%0.45 & -0.003 $\pm$ 0.011 &  0.198 $\pm$ 0.036 \\
%0.55 & 0.172 $\pm$ 0.016  &  0.535 $\pm$ 0.075 \\
%0.65 & 0.591 $\pm$ 0.034  &  0.986 $\pm$ 0.130 \\
%0.75 & 1.307 $\pm$ 0.063  &  1.413 $\pm$ 0.175 \\
\hline
\end{tabular}
\end{center}
\vskip 0.5cm

Contrary to~\cite{Alekhin03,ViMi}, we obtain different values of twist-four corrections for the hydrogen and deuterium data in NLO and NNLO. % (see Fig.~7).
Indeed, in the deuterium case, the NLO and NNLO corrections have the twist-four corrections insignificantly decreased, whereas in the hydrogen case the twist-four corrections are very small at NLO and NNLO.
It is quite reminiscent of an effect of HTC decreasing in NNLO observed earlier in~\cite{KKPS1} for $F_3$ SF.
%, but it seems that the reasons for this reduction in HTC are different. Here it is related with the cuts imposed on BCDMS data with large systematic errors that also considerably increases the coupling constant $\asMZ$, while
%the reason for HTC reduction in $F_3$ analysis comes from the small values of the higher twist terms for the latter at any order of perturbation theory~\footnote{For example, in~\cite{KoPe} the experimental data for $F_3$ SF were well described by only twist-two terms.}.

\vskip0.5cm
{\bf Table 7.} {\sl Parameter values of the twist-four term in different orders obtained in the analysis carried out within a fixed-flavor-number scheme ($n_f=4$) and no cut of BCDMS data with large systematics}
\begin{center}
\scriptsize
\begin{tabular}{|c||c|c||c|c||c|c|}
\hline
 & \multicolumn{2}{c||}{LO $\pm$ stat} & \multicolumn{2}{c||}{NLO $\pm$ stat} & \multicolumn{2}{c|}{NNLO $\pm$ stat}  \\
\cline{2-7}
$x$ & $\tilde h_4(x)$ for $H_2$ & $\tilde h_4(x)$ for $D_2$ & $\tilde h_4(x)$ for $H_2$ & $\tilde h_4(x)$ for $D_2$ & $\tilde h_4(x)$ for $H_2$ & $\tilde h_4(x)$ for $D_2$ \\
% & $\pm$ stat &  $\pm$ stat & $\pm$ stat &  $\pm$ stat & $\pm$ stat &  $\pm$ stat  \\
\hline \hline
0.275&-0.210$\pm$0.009&-0.193$\pm$0.015&-0.186$\pm$0.010&-0.176$\pm$0.011& -0.163 $\pm$ 0.010  & -0.155 $\pm$ 0.012\\
0.35 &-0.164$\pm$0.010&-0.110$\pm$0.021&-0.149$\pm$0.015&-0.106$\pm$0.015& -0.125 $\pm$ 0.010  & -0.087 $\pm$ 0.018 \\
0.45 & 0.026$\pm$0.016& 0.094$\pm$0.039& 0.009$\pm$0.031& 0.064$\pm$0.030&  0.020 $\pm$ 0.019  &  0.066 $\pm$ 0.040 \\
0.55 &0.337 $\pm$0.027& 0.478$\pm$0.067& 0.260$\pm$0.053&0.374 $\pm$0.050&  0.227 $\pm$ 0.031  &  0.324 $\pm$ 0.074 \\
0.65 &0.898 $\pm$0.058& 1.052$\pm$0.117& 0.719$\pm$0.092&0.827 $\pm$0.093&  0.590 $\pm$ 0.061  &  0.667 $\pm$ 0.130 \\
0.75 &1.508 $\pm$0.113& 1.256$\pm$0.178& 1.179$\pm$0.155&0.918 $\pm$0.134&  0.866 $\pm$ 0.115  &  0.606 $\pm$ 0.191 \\
%0.275& -0.163 $\pm$ 0.006 & -0.124 $\pm$ 0.010 \\
%0.35 & -0.133 $\pm$ 0.006 & -0.024 $\pm$ 0.016 \\
%0.45 & -0.003 $\pm$ 0.011 &  0.198 $\pm$ 0.036 \\
%0.55 & 0.172 $\pm$ 0.016  &  0.535 $\pm$ 0.075 \\
%0.65 & 0.591 $\pm$ 0.034  &  0.986 $\pm$ 0.130 \\
%0.75 & 1.307 $\pm$ 0.063  &  1.413 $\pm$ 0.175 \\
\hline
\end{tabular}
\end{center}

Note that the cut of the BCDMS data, which has increased the
$\as$ values (see Fig.~2) essentially improves agreement between perturbative QCD and
the experimental data.
Indeed, the HTCs that are nothing else but the difference between
the twist-two approximation (i.e. pure perturbative QCD contribution) and the
experimental data are seen to become considerably smaller at NLO and NNLO levels, to compare with NLO HT terms obtained in~\cite{ViMi} and also with the results of analysis obtained within a fixed-flavor-number scheme (with a number of flavors fixed to be 4) and no $Y$-cuts imposed on the BCDMS data(see Figs.~5,~6).

To make it clear with a HTC reduction effect, we perform a few more analyses:
\begin{itemize}
{\item within a fixed-flavor-number scheme (FFNS) and $n_f=4$ (i.e. no thresholds considered);}
{\item no $Y$-cuts imposed on BCDMS data with large systematic errors (i.e. with $N_{Y_{cut}}=0$);}
{\item the two above combined.}
\end{itemize}
As it is seen from Table~7 and Figs.~5 and 6, presented for the last case, without cuts and thresholds we reproduce the twist-four corrections obtained in~\cite{ViMi}.
The corresponding values of the coupling constant in NNLO are found to be
$$
\asMZ = 0.1082\quad\mathrm{for}\quad\chi^2_{SF} = 0.96\quad\mathrm{in\,\,the\,\,case\,\,of}\quad H_2\,\,\mathrm{data}\,,
$$
and
$$
\asMZ = 0.1094\quad\mathrm{for}\quad\chi^2_{SF} = 0.89\quad\mathrm{in\,\,the\,\,case\,\,of}\quad D_2\,\,\mathrm{data}\,.
$$
It looks like the effect induced by a particular choice of the threshold is small, however we plan to study different variants of heavy quark thresholds in our future investigations, the type of study carried out in, e.g., ~\cite{BlNee}.

\begin{figure}[!htb] %\label{HTC_comp}
\unitlength=1mm
\vskip -1.5cm
\begin{picture}(0,100)
\put(0,-5){\psfig{file=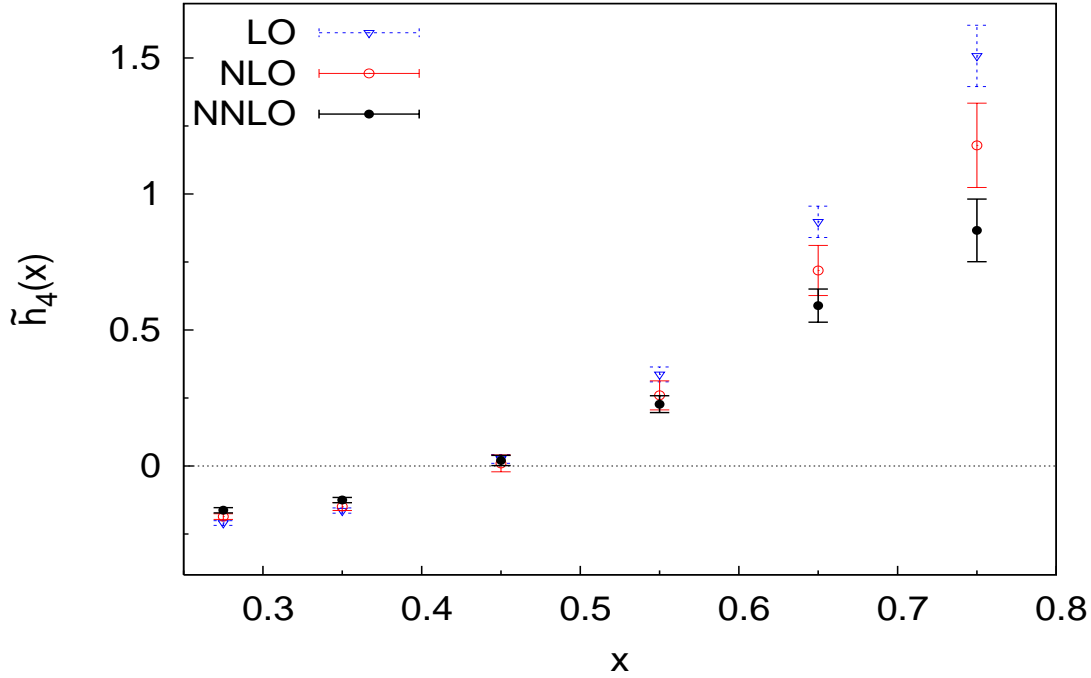,width=150mm,height=95mm}}
\end{picture}
\vskip 0.2cm
\caption{\it{Comparison of the HTC parameter $\tilde h_4(x)$ obtained at LO, NLO and NNLO for hydrogen data within a fixed-flavor-number scheme ($n_f=4$) and no $Y$ cuts imposed on the BCDMS data (i.e. the case $N_{Y_{cut}}=0$ in Tabl.~1)}.}
\end{figure}

\vskip 0.5cm
\begin{figure}[!htb] %\label{HTC_comp}
\unitlength=1mm
\vskip -1.5cm
\begin{picture}(0,100)
  \put(0,-5){%
   \psfig{file=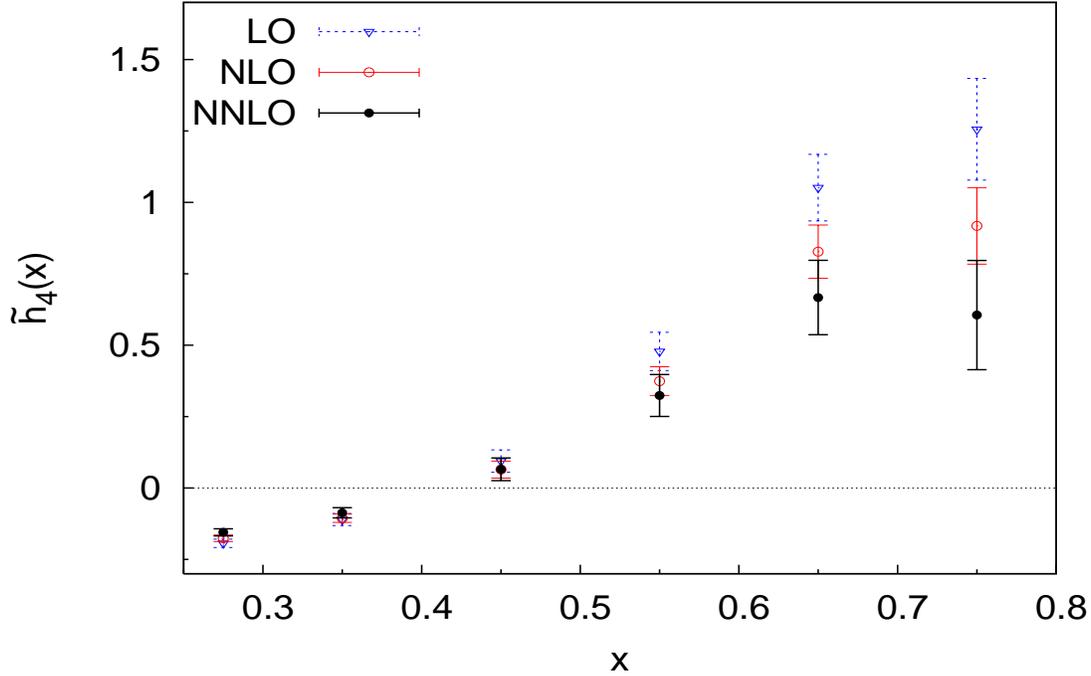,width=150mm,height=95mm}
}
\end{picture}
\vskip 0.2cm
\caption{\it{The same as in Fig.~5 in the case of deuterium data}.}
\end{figure}

Thus, the combined analysis of SLAC, NMC, BCDMS and BFP data in the
case of $Y$-cuts chosen corresponding to $N_{Y_{cut}}=6$ (see Table~1),
whenever HTC are not included and the $Q^2_{min}$ cut imposed is $8~\mathrm{GeV}^2$
(with a free normalization of the data sets), yields
(for $\chi^2\mbox{/DOF}~=~0.93$):
\bea
\as(90~\mbox{GeV}^2) &=& 0.1773 \pm 0.0018 ~\mbox{(stat)}\,, \nonumber \\
& &  \nonumber \\
\as(M_Z^2) &=& 0.1167 \pm 0.0008 ~\mbox{(stat)}\,,
\label{fu1.1}
\eea
and if HTC are included with the cut $Q^2 \geq 1$ GeV$^2$,
correspondingly ($\chi^2\mbox{/DOF}~=~0.98$):
\bea
\as(90~\mbox{GeV}^2) &=& 0.1772 \pm 0.0027 ~\mbox{(stat)}, \nonumber \\
& & \nonumber \\
\as(M_Z^2) &=& 0.1167 \pm 0.0010~\mbox{(stat)}\,.
\label{fu1.2}
\eea
It is seen that there is no substantial difference between the two,
therefore perturbative quantum chromodynamics seems to be applicable
with the cut $Q^2 \geq 8$ GeV$^2$ imposed as this nonsinglet analysis suggests.
%Therefore, it follows from the nonsinglet fits of experimental data carried
%out by far that perturbative QCD works rather well for $Q^2 \geq 7$ GeV$^2$.

Thus, using the analyses based on the nonsinglet evolution of the SLAC, NMC, BCDMS
and BFP experimental data for SF $F_2$ with no account for the twist-four corrections
and the cut $Q^2 \geq 8$ GeV$^2$ imposed, we obtain (for $\chi^2/DOF=0.93$)
\bea
\as(M_Z^2) ~=~ 0.1167 \pm 0.0008 ~\mbox{(stat)}
\pm 0.0018~\mbox{(syst)}  \pm 0.0007 ~\mbox{(norm)}
\label{NSfin}
\eea
or
\bea
\as(M_Z^2) ~=~ 0.1167 \pm 0.0021~\mbox{(total exp.error)}\,.
\label{NSfin1}
\eea

Upon including the twist-four corrections, and imposing the cut
$Q^2 \geq 1$ GeV$^2$, the following result is found
for $\chi^2/DOF=0.98$:
\bea
\as(M_Z^2) ~=~ 0.1167 \pm 0.0010 ~\mbox{(stat)}
\pm 0.0020 ~\mbox{(syst)}  \pm 0.0005 ~\mbox{(norm)}
\label{NSfin2}
\eea
or
\bea
\as(M_Z^2) ~=~ 0.1167 \pm 0.0022 ~\mbox{(total exp.error)}
\label{NSfin3}
\eea

%\vspace{0.5cm}
Looking at the results obtained in this section
one can note that similar to the NLO analysis the central value of the coupling
constant $\as(M_Z^2)$ obtained in the fits (NS evolution case) of the combined SLAC,
BCDMS, NMC and BFP data lie in-between the central values of the coupling constant obtained separately in the fits of BCDMS data alone and those of SLAC, NMC and BFP data analyzed together. Besides, all the values of $\as(M_Z^2)$ derived agree within existing statistical errors.

Within uncertainties, our result for $\as(M_Z^2)$ is also in good agreement with that cited in~\cite{NNLOBlumlein}
\be
\as(M_Z^2) = 0.1142 \pm 0.0023\,,
\label{NNLOBlumlein}
\ee
where a similar analysis of the NS part of the structure function $F_2$ has been performed.

%===================================================================
\section{Factorization and renormalization scale dependence}

In this section the dependence of the results on the different
choice of the factorization $\mu_F$ and renormalization $\mu_R$ scales
are examined. The threshold crossing point is taken to be at $Q^2_f=m^2_f$ because of its substantial role played in the evolution of the coupling constant~\cite{Shirkov97}.
Following the lines of the works~\cite{ViMi,scheme} we choose just three
values ($1/2,~1,~2$) for the coefficients $k_F$ and $k_R$.

Results are shown in Table~8. Fits are performed with no account for the higher twist corrections, with the number of points equal to $602$ (SLAC, BCDMS, NMC,
and BFP data), with $Q^2_{min} = 8$ GeV$^2$ and a free normalization for different data sets.
The change in the value of the coupling constant $\asMZ$ for various $k_F$
and $k_R$ values is denoted by the difference:
\bea
\Delta \as(M_Z^2) ~=~ \as(M_Z^2) - \as(M_Z^2)|_{k_F=k_R=1}\label{diff}
\eea

\newpage
{\bf Table 8.} $\asMZ$ {\sl for a set of} $k_F$ and $k_R$ {\sl coefficients}
\vspace{0.2cm}
\begin{center}
\begin{tabular}{|c|c||c|c|c|c|}
\hline
& & & & & \\
$k_R$ & $k_F$.  & $\chi^2(F_2)$ & $\as(90~\mbox{GeV}^2)$  $\pm$ stat &
$\asMZ$
%$\pm$ stat.err.
& $\Delta \as(M_Z^2)$ \\
& & & & & \\
\hline \hline
1 & 1 & 586 & 0.1773 $\pm$ 0.0018 & 0.1167 %$\pm$ 0.0012
& 0 \\
1/2 & 1 & 584 & 0.1734 $\pm$ 0.0017 & 0.1150 %$\pm$ 0.0020
& -0.0017 \\
1 & 1/2 & 585 & 0.1717 $\pm$ 0.0016 & 0.1143  %$\pm$ 0.0025
& -0.0024 \\
1 & 2 & 600 & 0.1845 $\pm$ 0.0021 & 0.1197 & +0.0030 \\
2 & 1 & 592 & 0.1829 $\pm$ 0.0020 & 0.1190 %\bf{0.1161}  %$\pm$ 0.0028
& +0.0023 \\
1/2 & 2 & 590 & 0.1795 $\pm$ 0.0019 & 0.1176 %$\pm$ 0.0011
& +0.0009 \\
2 & 1/2 & 584 & 0.1763 $\pm$ 0.0018 & 0.1163 %$\pm$ 0.0017
& -0.0004 \\
1/2 & 1/2 & 590 & 0.1689 $\pm$ 0.0015 & 0.1131  %$\pm$ 0.0011
& -0.0036 \\
2 & 2 & 609 & 0.1910 $\pm$ 0.0023 & 0.1223 %$\pm$ 0.0023
&  +0.0056\\
\hline
\end{tabular}
\end{center}
\vspace{0.5cm}

From Table~8 it follows that the theoretical uncertainties for the maximal and minimal values of the coupling constant that correspond to $k_R=k_F=2$ and $k_R=k_F=1/2$, respectively, are found to be $+0.0056$ and $-0.0036$, in order, thus reducing with respect to the NLO results obtained earlier~\cite{KK2001}.
It should be noted that we take into account the renormalization scale uncertainty in the expressions for the coefficient functions and the respective coupling constants analogously to what was done in~\cite{NeVo}.

Thus, using the analyses with NS evolution of the SLAC, NMC, BCDMS and BFP experimental data for SF $F_2$ we obtain for $\asMZ$ the following expressions (with no account for HTC, $Q^2 \geq 8$ GeV$^2$ and $\chi^2=0.93$):
\bea
\as(M_Z^2) &=& 0.1167 \pm 0.0008 ~\mbox{(stat)}
\pm 0.0018 ~\mbox{(syst)}  \pm 0.0007 ~\mbox{(norm)} \nonumber \\
&+&\biggl\{
\begin{array}{l}+ 0.0056 \\ -0.0036 \end{array} ~\mbox{(theor)},
\label{teo}
\eea
or
\bea
\as(M_Z^2) ~=~ 0.1167 \pm 0.0021 ~\mbox{(total exp.error)}
 +~ \biggl\{
\begin{array}{l} +0.0056 \\ -0.0036 \end{array} ~\mbox{(theor)}\,.
\label{teo1}
\eea

%=======================================================================
\section{Conclusions}

In this work the Jacobi polynomial expansion method developed in~\cite{Barker, Kri, Kri1}
was used to perform analysis of $Q^2$-evolution of DIS structure function $F_2$
by fitting all existing to date reliable fixed-target experimental data that satisfy the cut $x \geq 0.25$.
Based on the results of fitting the value of the QCD coupling constant at the normalization point was evaluated.
Starting with the reanalysis of BCDMS data by cutting off points with large systematic errors it was shown that the values of $\asMZ$ rise sharply with the cuts on systematics imposed.
On the other hand the latter do not depend on a certain cut within statistical errors.
The values $\asMZ$ obtained in various fits are in agreement with each other.
An outcome is that quite a similar result for $\asMZ$ was obtained in the analysis performed over BCDMS data (with the cuts on systematics) and over the data
of the rest, thus permitting us to fit available data altogether.

It turns out that for $Q^2 \geq 3$ GeV$^2$ the formul\ae\, of pure perturbative
QCD (i.e. twist-two approximation along with the target mass corrections)
are enough to achieve good agreement with all the data analyzed.
The reference result in NNLO is then found to be
\be
\as(M_Z^2) = 0.1167 \pm 0.0008 ~\mbox{(stat)}
\pm 0.0018 ~\mbox{(syst)} \pm 0.0007 ~\mbox{(norm)}, \label{re1n} \\
\ee

Upon adding twist-four corrections, fairly good agreement
between QCD (i.e. first two coefficients of Wilson expansion)
and the data starting already at $Q^2 = 1$ GeV$^2$, where the Wilson
expansion begins to be applicable, is observed.
This way we obtain for the coupling constant at $Z$ mass peak at NNLO level:
\be
\as(M_Z^2) = 0.1167 \pm 0.0007~\mbox{(stat)}
\pm 0.0020~\mbox{(syst)} \pm 0.0005~\mbox{(norm)}\,.
\label{re2n} \\
\ee

Note that there too is good agreement with the analysis~\cite{H1BCDMS} of the combined H1 and BCDMS data, which was published by H1 collaboration.
Our result for $\as(M_Z^2)$ is also compartible with the world average value for the coupling constant, presented in the review~\cite{Breview}~\footnote{It should be mentioned that this analysis was carried out over the data coming from the various experiments and in different orders of perturbation theory, i.e., from NLO up to N$^3$LO.}
$$
\as(M_Z^2) = 0.1184 \pm 0.0007\,,
%\label{world} \\
$$
or even more so if it is compared with the recent estimate given by MSTW group~\cite{GWatt}:
$$
\as(M_Z^2) = 0.1171 \pm 0.0014~(68\% C.L.) \pm 0.0034~(90\% C.L.)\,.
%\label{pdg} \\
$$

We would also like to note the importance of NNLO corrections in the analyses
of DIS experimental data. Incorporation of the NNLO corrections have been started already
several years ago in various ways. Results are based on the studies of higher order correction effects, which can be estimated from the dependence of our results on the factorization $\mu_F$ and renormalization $\mu_R$ scales.
As was pointed out the values of the theoretical uncertainties~\footnote{ As it has already been shown the scale choices $\mu_F=\mu_R=2Q^2$ and  $\mu_F=\mu_R=Q^2/2$ give the maximal and minimal values of $\as(M_Z^2)$ (at the various choices of values $k_F=1/2$, $k_F=2$, $k_R=1/2$ and $k_R=2$ separately) and thus give main part of theoretical error.},
given by this dependence of the results for $\as(M_Z^2)$ are equal to
\bea
\Delta\as(M_Z^2)|_{\mbox{theor}} ~=~
\biggl\{ \begin{array}{l} +0.0056 \nonumber \\ \nonumber
-0.0036 \end{array} \label{re3}\,.
\eea
For comparison let's quote the analogous numbers obtained at NLO~\cite{KK2001}:
$$
\Delta\as(M_Z^2)|_{\mbox{theor}} ~=~
\biggl\{ \begin{array}{l} +0.0070 \nonumber \\ \nonumber
-0.0041 \end{array}\,.
$$
Though the two cases cannot be directly compared, nonetheless some qualitative conclusions can be drawn. Thus, it is seen that the theoretical uncertainties stay still slightly higher than the total experimental error albeit somewhat less than those derived at NLO level. Perhaps, this calls for further account of even higher corrections (moreover, maybe the ones obtained within approaches different to that we stick with here) and is to be given elsewhere.
As it was shown in Refs.~\cite{NeVo,NeVo1}, the value of theoretical error should decrease approximately by a factor of 2 when the NNLO corrections are accounted for. This prediction is hardly observed, which can be attributed to a number of distinctions the two analyses bear in part.
Though a number of studies, devoted to NNLO QCD analysis of the structure functions and appeared in the literature (see~\cite{PKK}-\cite{KPS1},~\cite{SaYnd,NeVo,NeVo1,AlNNLO} and references therein) in the past, were exploiting back then partially known NNLO QCD corrections, it is obvious that in order to analyze experimental data across a whole
region of $x$ as precise as possible it is necessary to know all NNLO QCD corrections as exact as possible. These were evaluated in~\cite{MVV2004,MVV2005} and their exact expressions (rather than the approximate expressions given there as well) were used in this paper.

Concerning the contributions of higher twist corrections in the present work the well-known $x$-shape of the twist-four corrections while going from intermediate to large values of the Bjorken variable $x$ is well reproduced.
The latter look very similar to those from~\cite{ViMi}, if no cuts are imposed on BCDMS data with large systematic errors. The latter substantially reduce the twist-four corrections at NLO and NNLO level, particularly for hydrogen data.

The next step to take in the study is the consideration
of the combined nonsinglet and singlet analysis using the DIS experimental data
in the full $x$ region and also application of some resummation-like
Grunberg effective charge method~\cite{Grunberg} (as it was done in~\cite{Vovk} at the NLO approximation) and the ``frozen''~\cite{frozen}~\footnote{There are a lot of ``frozen'' versions of the strong coupling constant (see, for example, the list of references in \cite{Zotov}).} and analytic~\cite{SoShi} versions of the strong coupling constant (see~\cite{Zotov,ShiTer,CIKK09} for recent studies in this direction).

Moreover, we plan to consider also further corrections (i.e. those coming from three loops) in the coefficient functions~\cite{MVV2005}, which permits performing the N$^3$LO fits at large $x$ values, where the contributions of the corresponding four-loop corrections to the yet unknown anomalous dimensions should be negligible.
Several N$^3$LO fits had already been done in~\cite{KPS1,NNLOBlumlein,Iran}.
It will be carried out in nearest future with the purpose of studying further
reduction of theoretical uncertainties.

\section{Acknowledgments}
The work was supported by RFBR grant No.07-02-01046-a. The work of GP was supported by the grant Ministerio de Ciencia e Inovacion FPA2008-01177.

\end{document}